\documentclass[longauth,traditabstract]{aa}
\usepackage{graphicx}
\usepackage{amsmath,amsfonts,amssymb}
\usepackage{txfonts}
\usepackage[breaklinks,colorlinks,citecolor=blue,pdfa=true]{hyperref}
\usepackage{color}
\usepackage{fixltx2e}
\usepackage{natbib}
\usepackage{url}
\usepackage{multirow}
\usepackage{epsf}
\usepackage{epsfig}

\usepackage{ifthen}

\usepackage[T1]{fontenc}
\usepackage{lmodern}
\usepackage{ifxetex,ifluatex}

\usepackage{dblfloatfix}
\usepackage{morefloats}

\usepackage[switch]{lineno}

\bibpunct{(}{)}{;}{a}{}{,}

\def\setsymbol#1#2{\expandafter\def\csname #1\endcsname{#2}}
\def\getsymbol#1{\csname #1\endcsname}

\def\Planck{\textit{Planck}}

\newbox\tablebox    \newdimen\tablewidth
\def\leaderfil{\leaders\hbox to 5pt{\hss.\hss}\hfil}
\def\endPlancktable{\tablewidth=\columnwidth 
    $$\hss\copy\tablebox\hss$$
    \vskip-\lastskip\vskip -2pt}
\def\endPlancktablewide{\tablewidth=\textwidth 
    $$\hss\copy\tablebox\hss$$
    \vskip-\lastskip\vskip -2pt}
\def\tablenote#1 #2\par{\begingroup \parindent=0.8em
    \abovedisplayshortskip=0pt\belowdisplayshortskip=0pt
    \noindent
    $$\hss\vbox{\hsize\tablewidth \hangindent=\parindent \hangafter=1 \noindent
    \hbox to \parindent{$^#1$\hss}\strut#2\strut\par}\hss$$
    \endgroup}
\def\doubleline{\vskip 3pt\hrule \vskip 1.5pt \hrule \vskip 5pt}

\def\L2{\ifmmode L_2\else $L_2$\fi}

\def\DeltaT{\ifmmode \Delta T\else $\Delta T$\fi}
\def\deltat{\ifmmode \Delta t\else $\Delta t$\fi}
\def\fknee{\ifmmode f_{\rm knee}\else $f_{\rm knee}$\fi}
\def\Fmax{\ifmmode F_{\rm max}\else $F_{\rm max}$\fi}
\def\solar{\ifmmode{\rm M}_{\mathord\odot}\else${\rm M}_{\mathord\odot}$\fi}
\def\Msolar{\ifmmode{\rm M}_{\mathord\odot}\else${\rm M}_{\mathord\odot}$\fi}
\def\Lsolar{\ifmmode{\rm L}_{\mathord\odot}\else${\rm L}_{\mathord\odot}$\fi}
\def\inv{\ifmmode^{-1}\else$^{-1}$\fi}
\def\mo{\ifmmode^{-1}\else$^{-1}$\fi}
\def\sup#1{\ifmmode ^{\rm #1}\else $^{\rm #1}$\fi}
\def\expo#1{\ifmmode \times 10^{#1}\else $\times 10^{#1}$\fi}
\def\,{\thinspace}
\def\lsim{\mathrel{\raise .4ex\hbox{\rlap{$<$}\lower 1.2ex\hbox{$\sim$}}}}
\def\gsim{\mathrel{\raise .4ex\hbox{\rlap{$>$}\lower 1.2ex\hbox{$\sim$}}}}

\def\simprop{\mathrel{\raise .4ex\hbox{\rlap{$\propto$}\lower 1.2ex\hbox{$\sim$}}}}
\def\deg{\ifmmode^\circ\else$^\circ$\fi}
\def\pdeg{\ifmmode $\setbox0=\hbox{$^{\circ}$}\rlap{\hskip.11\wd0 .}$^{\circ}
          \else \setbox0=\hbox{$^{\circ}$}\rlap{\hskip.11\wd0 .}$^{\circ}$\fi}
\def\arcs{\ifmmode {^{\scriptstyle\prime\prime}}
          \else $^{\scriptstyle\prime\prime}$\fi}
\def\arcm{\ifmmode {^{\scriptstyle\prime}}
          \else $^{\scriptstyle\prime}$\fi}
\newdimen\sa  \newdimen\sb
\def\parcs{\sa=.07em \sb=.03em
     \ifmmode \hbox{\rlap{.}}^{\scriptstyle\prime\kern -\sb\prime}\hbox{\kern -\sa}
     \else \rlap{.}$^{\scriptstyle\prime\kern -\sb\prime}$\kern -\sa\fi}
\def\parcm{\sa=.08em \sb=.03em
     \ifmmode \hbox{\rlap{.}\kern\sa}^{\scriptstyle\prime}\hbox{\kern-\sb}
     \else \rlap{.}\kern\sa$^{\scriptstyle\prime}$\kern-\sb\fi}
\def\ra[#1 #2 #3.#4]{#1\sup{h}#2\sup{m}#3\sup{s}\llap.#4}
\def\dec[#1 #2 #3.#4]{#1\deg#2\arcm#3\arcs\llap.#4}
\def\deco[#1 #2 #3]{#1\deg#2\arcm#3\arcs}
\def\rra[#1 #2]{#1\sup{h}#2\sup{m}}

\def\dots{\relax\ifmmode \ldots\else $\ldots$\fi}
\def\WHzsr{\ifmmode $W\,Hz\mo\,sr\mo$\else W\,Hz\mo\,sr\mo\fi}
\def\mHz{\ifmmode $\,mHz$\else \,mHz\fi}
\def\GHz{\ifmmode $\,GHz$\else \,GHz\fi}
\def\mKs{\ifmmode $\,mK\,s$^{1/2}\else \,mK\,s$^{1/2}$\fi}
\def\muKs{\ifmmode \,\mu$K\,s$^{1/2}\else \,$\mu$K\,s$^{1/2}$\fi}
\def\muKRJs{\ifmmode \,\mu$K$_{\rm RJ}$\,s$^{1/2}\else \,$\mu$K$_{\rm RJ}$\,s$^{1/2}$\fi}
\def\muKHz{\ifmmode \,\mu$K\,Hz$^{-1/2}\else \,$\mu$K\,Hz$^{-1/2}$\fi}
\def\MJysr{\ifmmode \,$MJy\,sr\mo$\else \,MJy\,sr\mo\fi}
\def\MJysrmK{\ifmmode \,$MJy\,sr\mo$\,mK$_{\rm CMB}\mo\else \,MJy\,sr\mo\,mK$_{\rm CMB}\mo$\fi}
\def\microns{\ifmmode \,\mu$m$\else \,$\mu$m\fi}

\def\muK{\ifmmode \,\mu$K$\else \,$\mu$\hbox{K}\fi}
\def\microK{\ifmmode \,\mu$K$\else \,$\mu$\hbox{K}\fi}
\def\muW{\ifmmode \,\mu$W$\else \,$\mu$\hbox{W}\fi}
\def\kms{\ifmmode $\,km\,s$^{-1}\else \,km\,s$^{-1}$\fi}
\def\kmsMpc{\ifmmode $\,\kms\,Mpc\mo$\else \,\kms\,Mpc\mo\fi}
\providecommand{\sorthelp}[1]{}

\def\eqref#1{(\ref{#1})}

\def\smica{{\tt SMICA}}
\def\nilc{{\tt NILC}}
\def\sevem{{\tt SEVEM}}
\def\commander{\texttt{Commander}}

\def\eq{\begin{eqnarray}}
\def\qe{\end{eqnarray}}

\def\curl{\mathcal}
\def\({\left(}
\def\){\right)}

\def\and{\quad \mbox{and} \quad}
\def\barQ{\kern2pt\overline{\kern-2pt\curl{Q}}}

\def\barR{\kern2pt\overline{\kern-2pt\curl{R}}}

\def\bargamma{\kern2pt\overline{\kern-2pt\gamma}}

\def\leaderfil{\leaders\hbox to 5pt{\hss.\hss}\hfil}

\newcommand{\be}{\begin{equation}}
\newcommand{\ee}{\end{equation}}
\newcommand{\bea}{\begin{eqnarray}}
\newcommand{\eea}{\end{eqnarray}}

\renewcommand{\L}[0]{\mathbf{L}}

\def\inv{^{-1}}

\newcommand{\spinup}{\;\raise1.0pt\hbox{$'$}\hskip-6pt\partial\;}
\newcommand{\spindown}{\;\overline{\raise1.0pt\hbox{$'$}\hskip-6pt\partial}\;}

\graphicspath{{./figures/}}

\title{\Planck\ intermediate results. XLIX. Parity-violation constraints from polarization data}

\titlerunning{Parity violation constraints}
\authorrunning{Planck Collaboration}

\begin{document}

\author{\small
Planck Collaboration: N.~Aghanim\inst{47}
\and
M.~Ashdown\inst{57, 4}
\and
J.~Aumont\inst{47}
\and
C.~Baccigalupi\inst{67}
\and
M.~Ballardini\inst{23, 38, 41}
\and
A.~J.~Banday\inst{77, 7}
\and
R.~B.~Barreiro\inst{52}
\and
N.~Bartolo\inst{22, 53}
\and
S.~Basak\inst{67}
\and
K.~Benabed\inst{48, 76}
% ...
}
\institute{\small
APC, AstroParticule et Cosmologie, Universit\'{e} Paris Diderot, CNRS/IN2P3, CEA/lrfu, Observatoire de Paris, Sorbonne Paris Cit\'{e}, 10, rue Alice Domon et L\'{e}onie Duquet, 75205 Paris Cedex 13, France\goodbreak
\and
African Institute for Mathematical Sciences, 6-8 Melrose Road, Muizenberg, Cape Town, South Africa\goodbreak
\and
Agenzia Spaziale Italiana Science Data Center, Via del Politecnico snc, 00133, Roma, Italy\goodbreak
\and
Astrophysics Group, Cavendish Laboratory, University of Cambridge, J J Thomson Avenue, Cambridge CB3 0HE, U.K.\goodbreak
\and
Astrophysics \& Cosmology Research Unit, School of Mathematics, Statistics \& Computer Science, University of KwaZulu-Natal, Westville Campus, Private Bag X54001, Durban 4000, South Africa\goodbreak
\and
CITA, University of Toronto, 60 St. George St., Toronto, ON M5S 3H8, Canada\goodbreak
\and
CNRS, IRAP, 9 Av. colonel Roche, BP 44346, F-31028 Toulouse cedex 4, France\goodbreak
\and
California Institute of Technology, Pasadena, California, U.S.A.\goodbreak
\and
Computational Cosmology Center, Lawrence Berkeley National Laboratory, Berkeley, California, U.S.A.\goodbreak
\and
D\'{e}partement de Physique Th\'{e}orique, Universit\'{e} de Gen\`{e}ve, 24, Quai E. Ansermet,1211 Gen\`{e}ve 4, Switzerland\goodbreak
\and
Departamento de Astrof\'{i}sica, Universidad de La Laguna (ULL), E-38206 La Laguna, Tenerife, Spain\goodbreak
\and
Departamento de F\'{\i}sica, Universidad de Oviedo, Avda. Calvo Sotelo s/n, Oviedo, Spain\goodbreak
\and
Department of Astrophysics/IMAPP, Radboud University Nijmegen, P.O. Box 9010, 6500 GL Nijmegen, The Netherlands\goodbreak
\and
Department of Physics \& Astronomy, University of British Columbia, 6224 Agricultural Road, Vancouver, British Columbia, Canada\goodbreak
\and
Department of Physics and Astronomy, Dana and David Dornsife College of Letter, Arts and Sciences, University of Southern California, Los Angeles, CA 90089, U.S.A.\goodbreak
\and
Department of Physics and Astronomy, University College London, London WC1E 6BT, U.K.\goodbreak
\and
Department of Physics and Astronomy, University of Sussex, Brighton BN1 9QH, U.K.\goodbreak
\and
Department of Physics, Gustaf H\"{a}llstr\"{o}min katu 2a, University of Helsinki, Helsinki, Finland\goodbreak
\and
Department of Physics, Princeton University, Princeton, New Jersey, U.S.A.\goodbreak
\and
Department of Physics, University of California, Santa Barbara, California, U.S.A.\goodbreak
\and
Department of Physics, University of Illinois at Urbana-Champaign, 1110 West Green Street, Urbana, Illinois, U.S.A.\goodbreak
\and
Dipartimento di Fisica e Astronomia G. Galilei, Universit\`{a} degli Studi di Padova, via Marzolo 8, 35131 Padova, Italy\goodbreak
\and
Dipartimento di Fisica e Astronomia, Alma Mater Studiorum, Universit\`{a} degli Studi di Bologna, Viale Berti Pichat 6/2, I-40127, Bologna, Italy\goodbreak
\and
Dipartimento di Fisica e Scienze della Terra, Universit\`{a} di Ferrara, Via Saragat 1, 44122 Ferrara, Italy\goodbreak
\and
Dipartimento di Fisica, Universit\`{a} La Sapienza, P. le A. Moro 2, Roma, Italy\goodbreak
\and
Dipartimento di Fisica, Universit\`{a} degli Studi di Milano, Via Celoria, 16, Milano, Italy\goodbreak
\and
Dipartimento di Fisica, Universit\`{a} degli Studi di Trieste, via A. Valerio 2, Trieste, Italy\goodbreak
\and
Dipartimento di Fisica, Universit\`{a} di Roma Tor Vergata, Via della Ricerca Scientifica, 1, Roma, Italy\goodbreak
\and
Dipartimento di Matematica, Universit\`{a} di Roma Tor Vergata, Via della Ricerca Scientifica, 1, Roma, Italy\goodbreak
\and
European Space Agency, ESAC, Planck Science Office, Camino bajo del Castillo, s/n, Urbanizaci\'{o}n Villafranca del Castillo, Villanueva de la Ca\~{n}ada, Madrid, Spain\goodbreak
\and
European Space Agency, ESTEC, Keplerlaan 1, 2201 AZ Noordwijk, The Netherlands\goodbreak
\and
Gran Sasso Science Institute, INFN, viale F. Crispi 7, 67100 L'Aquila, Italy\goodbreak
\and
HGSFP and University of Heidelberg, Theoretical Physics Department, Philosophenweg 16, 69120, Heidelberg, Germany\goodbreak
\and
Helsinki Institute of Physics, Gustaf H\"{a}llstr\"{o}min katu 2, University of Helsinki, Helsinki, Finland\goodbreak
\and
INAF - Osservatorio Astronomico di Padova, Vicolo dell'Osservatorio 5, Padova, Italy\goodbreak
\and
INAF - Osservatorio Astronomico di Roma, via di Frascati 33, Monte Porzio Catone, Italy\goodbreak
\and
INAF - Osservatorio Astronomico di Trieste, Via G.B. Tiepolo 11, Trieste, Italy\goodbreak
\and
INAF/IASF Bologna, Via Gobetti 101, Bologna, Italy\goodbreak
\and
INAF/IASF Milano, Via E. Bassini 15, Milano, Italy\goodbreak
\and
INFN - CNAF, viale Berti Pichat 6/2, 40127 Bologna, Italy\goodbreak
\and
INFN, Sezione di Bologna, viale Berti Pichat 6/2, 40127 Bologna, Italy\goodbreak
\and
INFN, Sezione di Ferrara, Via Saragat 1, 44122 Ferrara, Italy\goodbreak
\and
INFN, Sezione di Roma 1, Universit\`{a} di Roma Sapienza, Piazzale Aldo Moro 2, 00185, Roma, Italy\goodbreak
\and
INFN, Sezione di Roma 2, Universit\`{a} di Roma Tor Vergata, Via della Ricerca Scientifica, 1, Roma, Italy\goodbreak
\and
IPAG: Institut de Plan\'{e}tologie et d'Astrophysique de Grenoble, Universit\'{e} Grenoble Alpes, IPAG, F-38000 Grenoble, France, CNRS, IPAG, F-38000 Grenoble, France\goodbreak
\and
Imperial College London, Astrophysics group, Blackett Laboratory, Prince Consort Road, London, SW7 2AZ, U.K.\goodbreak
\and
Institut d'Astrophysique Spatiale, CNRS, Univ. Paris-Sud, Universit\'{e} Paris-Saclay, B\^{a}t. 121, 91405 Orsay cedex, France\goodbreak
\and
Institut d'Astrophysique de Paris, CNRS (UMR7095), 98 bis Boulevard Arago, F-75014, Paris, France\goodbreak
\and
Institute of Astronomy, University of Cambridge, Madingley Road, Cambridge CB3 0HA, U.K.\goodbreak
\and
Institute of Theoretical Astrophysics, University of Oslo, Blindern, Oslo, Norway\goodbreak
\and
Instituto de Astrof\'{\i}sica de Canarias, C/V\'{\i}a L\'{a}ctea s/n, La Laguna, Tenerife, Spain\goodbreak
\and
Instituto de F\'{\i}sica de Cantabria (CSIC-Universidad de Cantabria), Avda. de los Castros s/n, Santander, Spain\goodbreak
\and
Istituto Nazionale di Fisica Nucleare, Sezione di Padova, via Marzolo 8, I-35131 Padova, Italy\goodbreak
\and
Jet Propulsion Laboratory, California Institute of Technology, 4800 Oak Grove Drive, Pasadena, California, U.S.A.\goodbreak
\and
Jodrell Bank Centre for Astrophysics, Alan Turing Building, School of Physics and Astronomy, The University of Manchester, Oxford Road, Manchester, M13 9PL, U.K.\goodbreak
\and
Kavli Institute for Cosmological Physics, University of Chicago, Chicago, IL 60637, USA\goodbreak
\and
Kavli Institute for Cosmology Cambridge, Madingley Road, Cambridge, CB3 0HA, U.K.\goodbreak
\and
LAL, Universit\'{e} Paris-Sud, CNRS/IN2P3, Orsay, France\goodbreak
\and
LERMA, CNRS, Observatoire de Paris, 61 Avenue de l'Observatoire, Paris, France\goodbreak
\and
Laboratoire Traitement et Communication de l'Information, CNRS (UMR 5141) and T\'{e}l\'{e}com ParisTech, 46 rue Barrault F-75634 Paris Cedex 13, France\goodbreak
\and
Laboratoire de Physique Subatomique et Cosmologie, Universit\'{e} Grenoble-Alpes, CNRS/IN2P3, 53, rue des Martyrs, 38026 Grenoble Cedex, France\goodbreak
\and
Laboratoire de Physique Th\'{e}orique, Universit\'{e} Paris-Sud 11 \& CNRS, B\^{a}timent 210, 91405 Orsay, France\goodbreak
\and
Max-Planck-Institut f\"{u}r Astrophysik, Karl-Schwarzschild-Str. 1, 85741 Garching, Germany\goodbreak
\and
Mullard Space Science Laboratory, University College London, Surrey RH5 6NT, U.K.\goodbreak
\and
Nicolaus Copernicus Astronomical Center, Bartycka 18, 00-716 Warsaw, Poland\goodbreak
\and
Nordita (Nordic Institute for Theoretical Physics), Roslagstullsbacken 23, SE-106 91 Stockholm, Sweden\goodbreak
\and
SISSA, Astrophysics Sector, via Bonomea 265, 34136, Trieste, Italy\goodbreak
\and
School of Chemistry and Physics, University of KwaZulu-Natal, Westville Campus, Private Bag X54001, Durban, 4000, South Africa\goodbreak
\and
School of Physics and Astronomy, Cardiff University, Queens Buildings, The Parade, Cardiff, CF24 3AA, U.K.\goodbreak
\and
School of Physics and Astronomy, University of Nottingham, Nottingham NG7 2RD, U.K.\goodbreak
\and
Simon Fraser University, Department of Physics, 8888 University Drive, Burnaby BC, Canada\goodbreak
\and
Sorbonne Universit\'{e}-UPMC, UMR7095, Institut d'Astrophysique de Paris, 98 bis Boulevard Arago, F-75014, Paris, France\goodbreak
\and
Space Sciences Laboratory, University of California, Berkeley, California, U.S.A.\goodbreak
\and
Sub-Department of Astrophysics, University of Oxford, Keble Road, Oxford OX1 3RH, U.K.\goodbreak
\and
The Oskar Klein Centre for Cosmoparticle Physics, Department of Physics,Stockholm University, AlbaNova, SE-106 91 Stockholm, Sweden\goodbreak
\and
UPMC Univ Paris 06, UMR7095, 98 bis Boulevard Arago, F-75014, Paris, France\goodbreak
\and
Universit\'{e} de Toulouse, UPS-OMP, IRAP, F-31028 Toulouse cedex 4, France\goodbreak
% ...
}

\abstract{
Parity violating extensions of the standard electromagnetic theory cause in
vacuo rotation of the plane of polarization of propagating photons.
This effect, also known as cosmic birefringence, impacts the cosmic microwave
background (CMB) anisotropy angular power spectra, producing non-vanishing $T$--$B$ and $E$--$B$ correlations
that are otherwise null when parity is a symmetry.
Here we present new constraints on an isotropic rotation, parametrized by the
angle $\alpha$, derived from  \Planck\ 2015 CMB polarization data.
To increase the robustness of our analyses,  we employ two complementary
approaches, in harmonic space and in map space, the latter based on a peak stacking technique.
The two approaches provide estimates for $\alpha$ that are in agreement within
statistical uncertainties and very stable against several consistency tests.
Considering the $T$--$B$ and $E$--$B$ information jointly, we find $\alpha = 0\fdg31 \pm
0\fdg05\, ({\rm stat.})\, \pm 0\fdg28\, ({\rm syst.})$ from the harmonic
analysis and $\alpha = 0\fdg35 \pm 0\fdg05\, ({\rm stat.})\, \pm 0\fdg28\, ({\rm
syst.})$ from the stacking approach.
These constraints are compatible with no parity violation and are dominated by the systematic
uncertainty in the orientation of \Planck's polarization-sensitive bolometers.}

\maketitle

\section{Introduction}
\label{sec_introduction}

Measuring the in vacuo rotation of the plane of polarization of photons is a way to test fundamental
physics in the Universe. Such a rotation is sensitive to parity-violating
interactions in the electromagnetic sector that are found in extensions of the Standard
Model of particle physics \citep[][]{Carroll1998,Lue1999,Feng2005,Li2009}.
For example, extending the Maxwell Lagrangian with a coupling (scalar,
Chern-Simons, etc.) to
$A_{\nu}\tilde{F}^{\mu\nu}$,\footnote{Here $A_\nu$ is the photon field, and
$\tilde{F}^{\mu\nu}$ is the dual of the Faraday tensor, defined
to be $\tilde{F}^{\mu\nu} \equiv (1/2) \epsilon^{\mu\nu\rho\sigma}
F_{\rho\sigma}$.} impacts right- and left-handed photons asymmetrically.
Therefore a photon at the last-scattering surface with linear polarization in one
orientation will arrive at our detectors with its plane of polarization
rotated due to this coupling term. The amount of rotation, usually denoted $\alpha$,
is often referred to as the cosmic birefringence angle. This rotation naturally
mixes $E$- and $B$-modes of CMB polarization\footnote{We use the customary convention used by the CMB community
for the Q and U Stokes parameters, see e.g.
\url{http://wiki.cosmos.esa.int/planckpla2015/index.php/Sky\_temperature\_maps}.} and generates $T$--$B$ and
$E$--$B$ correlations that would be zero in the absence of parity violations.
The cosmic microwave background (CMB) polarization is particularly useful for
measuring such an effect, because even if the coupling is small, CMB photons have
travelled a large comoving distance from the last-scattering surface (almost)
completely unimpeded and thus the rotation could accumulate into a measurable
signal.

This effect has previously been investigated using data from many CMB
experiments \citep[][]{Feng2006, Wu2009, Brown2009, Pagano2009, Komatsu2011,Hinshaw:2012aka,
Ade:2014afa, Ade:2014gua, Kaufman2014, Naess2014, Alighieri:2014yoa, Zhao:2015mqa, Mei2015, Gruppuso2015, Contaldi2015, Molinari:2016xsy}, and also by looking at
radio galaxy data \citep[][]{Carroll1990, Cimatti1993, Cimatti1994, Wardle1997,
Leahy1997, Carroll1998, diSeregoAlighieri2010, Kamionkowski2010}.  Thus far all
the constraints are compatible with no cosmic birefringence (see
discussion in Sect.~\ref{sec:conclusions}).

In this paper we employ \Planck\footnote{\Planck\
  (\url{http://www.esa.int/Planck}) is a project of the European Space Agency
  (ESA) with instruments provided by two scientific consortia funded by ESA
  member states and led by Principal Investigators from France and Italy,
  telescope reflectors provided through a collaboration between ESA and a
  scientific consortium led and funded by Denmark, and additional contributions
  from NASA (USA).} 2015 CMB data to estimate an isotropic $\alpha$.
The birefringence angle has already been constrained with \Planck\ data in
\citet{Gruppuso2015}, using the publicly available 2015 \Planck\ Likelihood \citep[][]{planck2014-a13}.
However, that work did not use $T$--$B$ and $E$--$B$ data, which
are essential for determining the sign of $\alpha$
and for increasing the constraining power. We include here $T$--$B$ and $E$--$B$ cross-correlations by considering two approaches,
one based on harmonic space, through the use of the so-called
``$D$-estimators'' and one
based on pixel-space maps that employs stacked images of the (transformed) $Q_{\rm
r}$ and $U_{\rm r}$
Stokes parameters.

The paper is organized as follows. In Sect. \ref{impact} we describe the
effect that cosmological birefringence has on the angular power spectra of the CMB.
In Sect. \ref{sec_dataset_and_simulations} we provide details of the data and
simulations that are considered in our analysis, which is described in Sect.
\ref{sec_analysis}.  Results for our two different methodologies are summarized
and compared
in Sect. \ref{sec_results}. Section \ref{sec:systematics} contains a discussion of the
systematic effects that are most important for the observables considered.
Finally,
conclusions are drawn in Sect. \ref{sec:conclusions}.

\section{Impact of birefringence on the CMB polarization spectra}
\label{impact}

Birefringence rotates the six CMB angular power spectra in the following way \citep[see][for more details]{Lue1999,Feng2006}:
\begin{align}
  C'^{TT}_{\ell} &= C^{TT}_{\ell};
  \label{eq:tt}\\
  C'^{EE}_{\ell} &= C^{EE}_{\ell} \cos^2{(2\alpha)} + C^{BB}_{\ell}
  \sin^2{(2\alpha)};
  \label{eq:ee}\\
  C'^{BB}_{\ell} &= C^{EE}_{\ell} \sin^2{(2\alpha)} + C^{BB}_{\ell}
  \cos^2{(2\alpha)};
  \label{eq:bb}\\
  C'^{TE}_{\ell} &= C^{TE}_{\ell} \cos{(2\alpha)};
  \label{eq:te}\\
  C'^{TB}_{\ell} &= C^{TE}_{\ell} \sin{(2\alpha)};
  \label{eq:tb}\\
  C'^{EB}_{\ell} &= \frac{1}{2} (C^{EE}_{\ell} - C^{BB}_{\ell})
  \sin{(4\alpha)} .
  \label{eq:eb}
\end{align}
Here $\alpha$ is assumed to be constant (see \citealt{Liu:2006uh,Finelli:2008jv,Li:2008tma} for generalizations).
In this paper we will consider only the above parametrization, where the
primed $C'_{\ell}$ are the observed spectra and the unprimed $C_{\ell}$ are the
spectra one would measure in the absence of parity violations.
In principle the rotation angle $\alpha$ could depend on direction (with details dictated by the specific model considered),
and one could measure the anisotropies of $\alpha$. We do not employ this type of
analysis here, but focus on the simple case of an {\it isotropic} $\alpha$ (or the
$\alpha$ monopole) (see \citealt{Gluscevic:2012me} and \citealt{Ade:2015cao} for
constraints on anisotropic birefringence).

Isotropic birefringence is indistinguishable from a systematic, unknown
mismatch of the global orientation of the polarimeters.
This is strictly true if the cosmological birefringence $\alpha$ is the same regardless of the multipole $\ell$ at which CMB polarization is measured.
However, specific birefringence models may predict some angular dependence in $\alpha$.
Furthermore,
large angular scale polarization in $\alpha$ is sourced in the re-ionization epoch, as opposed to
the small scales which are formed at recombination \citep[][]{Komatsu2011,Gruppuso2015}.
This will inevitably produce some angular dependency in $\alpha$ (assuming that
the birefringence angle is proportional to the CMB photon path)
and this effect could in principle be used to disentangle instrumental
systematic effects (since photons that scattered at the re-ionization epoch would have traveled less than the others).
However, we focus here on smaller scale data, where the reionization effects
are not important and therefore such a distinction is not possible.
For \Planck\ there is an estimate of the uncertainty of the possible instrument
polarization angle using measurements performed on the ground
\citep{Rosset2010}, as discussed further in Sect.~\ref{sec:systematics}.
Unfortunately, in-flight calibration is complicated by the scarcity of
linearly polarized sources that are bright enough, with the Crab Nebula being a
primary calibration source \citep{planck2014-a09}.\footnote{We do not employ
any self-calibration procedure as suggested in \citet{Keating:2012ge}, since it is degenerate
with the effect we are looking for.}

Eqs.~(\ref{eq:tt})--(\ref{eq:eb}) include all the secondary anisotropies but the
weak-lensing effect. Due the current precision of data \citep[see the
discussion in][]{Gubitosi2014} we safely ignore the weak-lensing effect as it
contributes a negligible error.

\section{Data and simulations}
\label{sec_dataset_and_simulations}

We use the full-mission \Planck\ \citep{planck2014-a01} component-separated
temperature and high-pass-filtered polarization maps at {\tt
HEALPix}\footnote{\url{http://healpix.sourceforge.net/}}
\citep[][]{Gorski2005} resolution $N_{\rm side} = 1024$; i.e., we take the
\commander, \nilc, \sevem, and \smica\ solutions for $T$, $Q$, $U$, and $E$,
  fully described in \citet{planck2014-a11} and \citet{planck2014-a12}, and available on the Planck Legacy
  Archive.\footnote{\url{http://www.cosmos.esa.int/web/planck/pla}}
The $E$-mode maps are calculated using the method of \citet[see also
\citealt{Kim2011}]{Bielewicz2012}.
We use the common temperature and polarization
masks at $N_{\rm side} = 1024$, namely {\tt UT$_\mathrm{1024}$76} and {\tt
UPB77},
respectively. For the harmonic analysis we also use half-mission data provided
by the \smica\ component-separation pipeline, in order to build our
$D^{EB}$-estimator from cross-correlations
\citep[][]{planck2014-a11}.
No further
smoothing is applied to any of the maps (although this version of the data
already includes 10\arcm\ smoothing in both temperature and polarization).

We note that there are known systematic effects in the polarization maps
released by \Planck\ that have not been fully remedied in the 2015 release
(see Sect.~\ref{sec:systematics} for a full discussion on the main systematic
effects relevant for this analysis).  These issues include various sources of
large angular scale artefacts, temperature-to-polarization leakage
\citep{planck2014-a08,planck2014-a09,planck2014-a13}, and a mismatch in noise
properties between the data and simulations \citep{planck2014-a14}. In order to
mitigate any large-angle artefacts, we use
only the high-pass-filtered version of the polarization data. We note that
neglecting the large scales
has little to no impact on our constraining power for $\alpha$.  We have also checked that
temperature-to-polarization leakage \citep[][]{planck2014-a13} has very little
effect on our analysis (see Sect.~\ref{sec:beams}); similar conclusions are
reached in \citet{planck2014-a10}.

We pay particular attention to the mis-characterization of the noise in the
polarization data. Given the recommendation in \citet{planck2014-a11} we
restrict our analysis to cross-correlation and stacking methods, which
are less sensitive to such noise issues \citep{planck2014-a09}.
For the harmonic analysis, we
estimate the angular power spectra up to multipoles $\ell \simeq 1500$
\citep[as suggested by the cosmological analysis tests carried out
in][]{planck2014-a11}, using simulations to create a $\chi^2$ statistic.
The map-space analysis does not require the use of simulations,\footnote{We explicitly
checked that using simulations does not change the results, which is simply a
consequence of the fact
that the process of stacking means we are not very sensitive to the
noise properties of the data.} since
we only need a relatively crude noise estimate on the scales we work at and we use a
weighting approach when stacking that is only dependent on the data (we have also
checked that our results are quite insensitive to the noise level of the data,
see Sect.~\ref{sec:pnoise}).
Nevertheless it is reassuring that the map-space and harmonic-space analyses
 arrive at consistent results.

We use realistic full focal plane (FFP8.1) simulations described in detail in
\citet{planck2014-a14}. These are $10^3$ simulations processed through the four
\Planck\
component-separation pipelines, namely \commander, \nilc, \sevem, and \smica\
\citep{planck2014-a11},  using the same weights as derived from the
\Planck\ full
mission data. The CMB output maps are used to build the harmonic space estimators
used in this work. For our harmonic space $EB$ estimator we use the half-mission
simulations provided by the \smica\ pipeline.

The FFP8.1 fiducial cosmology corresponds to the cosmological
parameters $\omega_{\rm b} = 0.0222$, $\omega_{\rm c} = 0.1203$, $\omega_\nu = 0.00064$,
$\Omega_\Lambda = 0.6823$, $h = 0.6712$, $n_{\rm s} = 0.96$, $A_{\rm s} = 2.09\times
10^{-9}$, and $\tau = 0.065$ (where $\omega_{\rm x} \equiv \Omega_{\rm x}
h^2$). Note that we perform the analysis for the birefringence angle by fixing the other cosmological parameters
to the values reported above. This seems to be a safe assumption,
since in \citet{Gruppuso2015} it was shown that $\alpha$ is quite
decoupled from the other parameters,
at least as long as $C_{\ell}^{TT}$, $C_{\ell}^{TE}$, and $C_{\ell}^{EE}$ are considered;
$\Lambda$CDM parameters are not expected to be constrainted much from
$C_{\ell}^{TB}$ and $C_{\ell}^{EB}$,
contrary to models that explicitly break parity symmetry.

\section{Analysis}
\label{sec_analysis}

\subsection{Map-space analysis}

We follow the stacking approach first introduced in \citet{Komatsu2011},
where they were able to constrain $\alpha$ by stacking polarization on temperature
extrema. Here we perform the same analysis, but also stack on $E$-mode extrema.
Our analysis is performed in map space (although we must briefly go to harmonic
space for stacking on $E$-modes, as described in Sect.~\ref{sec:pnoise}) and
we show that stacking polarization on temperature extrema is sensitive to the
$T$--$E$ and $T$--$B$ correlations, while stacking on $E$-mode extrema is
sensitive to the $E$--$E$ and $E$--$B$ correlations.

The recommendation on the use of polarization data from \citet{planck2014-a11}
is that only results with weak dependence on noise are to be considered
completely reliable. For the
purposes of stacking on temperature peaks only cross-correlation information is
used, and thus understanding the detailed noise properties of polarization is
unnecessary. Stacking on $E$-mode peaks the results {\it
do} depend on the noise properties of the map; this is because the expected
angular profiles of the stacks depend on the full power spectrum of the map. In
Sect.~\ref{sec:pnoise} we demonstrate that even a strong miscalculation of the noise
would result in shifts at below the $1\,\sigma$ level (and more reasonable
miscalculations of the noise will bias results at an essentially negligible level).

\subsubsection{$Q_{\rm r}$ and $U_{\rm r}$ parameters}

We use the transformed Stokes parameters $Q_{\rm r}$, and $U_{\rm r}$, first introduced in
\citet{Kamionkowski1997}:
\begin{align}
  Q_\mathrm{r}(\vec{\theta}) &= -Q(\vec{\theta})\cos{(2\phi)}
  -U(\vec{\theta})\sin{(2\phi)}; \label{eq:def_qr} \\
  U_\mathrm{r}(\vec{\theta}) &= Q(\vec{\theta})\sin{(2\phi)}
  -U(\vec{\theta})\cos{(2\phi)}.  \label{eq:def_ur}
\end{align}
Here $\phi$ is defined as the angle from a local ``east'' (where ``north''
always points towards the Galactic north pole) direction in the
coordinate system defined by centring on the hot or cold spot, and
$\vec{\theta}$ is a radial vector. The stacking procedure tends to produce
images with azimuthal symmetry, and hence the predictions will only depend on
$\theta$. The theoretical angular profiles for stacking on temperature hot
spots are derived in \citet[see also \citealt{planck2014-a18}]{Komatsu2011} and are explicitly given by
\begin{align}
 \left\langle Q^T_r \right\rangle (\theta) &=
  - \int \frac{\ell d\ell}{2\pi} W^T_{\ell} W^P_{\ell}
  \left(\bar{b}_\nu + \bar{b}_\zeta \ell^2\right) C^{TE}_{\ell} J_2(\ell\theta),
  \label{eq:qrprofile}\\
 \left\langle U^T_r \right\rangle (\theta) &= - \int \frac{\ell d\ell}{2\pi}
  W^T_{\ell} W^P_{\ell} \left(\bar{b}_\nu + \bar{b}_\zeta {\ell}^2\right)
  C^{TB}_{\ell} J_2({\ell}\theta). \label{eq:urprofile}
\end{align}
The quantities $W^{T,P}_{\ell}$ are combinations of the beam (10\arcm\
smoothing) and pixel window functions (at $N_{\rm{side}}=1024$) for temperature
and polarization. Below we will use $W^E_{\ell}$ to denote the same
quantity for $E$-modes; however, the $E$-modes are produced at the same
resolution as temperature and so $W^{E}_{\ell} = W^T_{\ell}$. The bracketed
term in each of Eqs.~\eqref{eq:qrprofile} and \eqref{eq:urprofile} incorporates the scale-dependent bias when converting the
underlying density field to temperature or $E$-modes (thus, they will differ if
the stacking is performed on temperature or $E$-mode extrema). The function $J_2$ is the
second-order Bessel function of the first kind. Angular profiles
derived from stacking on $E$-mode hot spots can easily be generalized from the
above formulae by simply noting that $E$-modes share the same statistical
properties as
temperature and thus we only need to change the power spectra in the above
formulae. Thus the angular profiles for stacking on $E$-mode hot spots are
given by
\begin{align}
 \left\langle Q^E_{\rm r} \right\rangle (\theta) =
  - \int \frac{\ell d\ell}{2\pi} W^E_{\ell} W^P_{\ell}&
  \left(\bar{b}_\nu + \bar{b}_\zeta \ell^2\right) \notag\\
  & \left(C^{EE}_{\ell} +
  N^{EE}_{\ell}\right) J_2(\ell\theta),
  \label{eq:Eqrprofile}\\
 \left\langle U^E_{\rm r} \right\rangle (\theta) = - \int \frac{\ell d\ell}{2\pi}
  W^E_{\ell} W^P_{\ell}& \left(\bar{b}_\nu + \bar{b}_\zeta {\ell}^2\right)
  C^{EB}_{\ell} J_2({\ell}\theta). \label{eq:Eurprofile}
\end{align}
The specific forms of $b_{\nu}$ (the scale-independent part) and $b_{\zeta}$
(which is proportional to second derivatives that define the peak) are
given in \citet{Desjacques2008}. The $\Lambda$CDM prediction for
$\left\langle U^{T,E}_{\rm r} \right\rangle$ is identically zero and thus we will
find that the vast majority of the constraining power comes from these
profiles. We also show explicitly in
Eqs.~\eqref{eq:qrprofile}--\eqref{eq:Eurprofile} that $Q_{\rm r}$ and $U_{\rm
r}$ are
sensitive to the $T$--$E$ and $T$--$B$ correlation when stacking on temperature
extrema or the $E$--$E$ and $E$--$B$ correlation when stacking on $E$-mode
extrema. Determination of the bias parameters depends on the power
spectrum of the map where the extrema are determined \citep[see][and
Appendix~\ref{app:Emodes}]{Komatsu2011}, thus they depend on the noise
properties of the map, as well as the underlying power spectrum.
Section~\ref{sec:pnoise} will examine to what extent misunderstanding the
noise might bias the results.

For our main results we have selected extrema using a threshold of $\nu = 0$,
which mean we consider {\it all\/} positive hot spots (or negative cold spots);
however, we have checked other choices of threshold and found consistency,
provided that we do not choose such a high a threshold such that the overall
signal becomes too weak. We do not claim that our analysis is optimal, and it
may be that a better weighting exists for different levels of threshold;
however, tests have shown that, in terms of minimizing the uncertainty on
$\alpha$, the choice of $\nu=0$ and use of averaged bias parameters is close to
optimal.

For the \Planck\ temperature data we calculate the bias parameters to be
$\bar{b}_{\nu} = 3.829\times10^{-3}\,\mu$K$^{-1}$ and $\bar{b}_{\zeta} =
1.049\times10^{-7}\,\mu$K$^{-1}$. For the \Planck\ $E$-mode data we calculate
$\bar{b}_{\nu} = (3.622,\, 3.384,\, 2.957,\,
3.332)\times10^{-2}\,\mu$K$^{-1}$ and $\bar{b}_{\zeta} = (1.727,\, 3.036,\,
1.874,\, 3.039)\times10^{-7}\,\mu$K$^{-1}$ for \commander, \nilc, \sevem, and
\smica, respectively. The derivation of
Eqs.~\eqref{eq:qrprofile}--\eqref{eq:urprofile} and a discussion of how to
calculate all relevant quantities are given in appendix~B of
\citet{Komatsu2011}, while the derivation of
Eqs.~\eqref{eq:Eqrprofile}--\eqref{eq:Eurprofile} is given in
Appendix~\ref{app:Emodes} of this paper. The reader is referred to
\citet{Komatsu2011} and also section 8 of of \citet{planck2014-a18} for a complete description
of the physics behind the features in the predicted stacked profiles.

\subsubsection{Procedure}

We begin by locating all local extrema\footnote{As previously mentioned, we use
{\it all\/} positive (negative) local maxima (minima) for hot (cold) spots. Extrema are
defined by comparing each pixel to its nearest neighbours.} of the temperature
(or $E$-mode) data outside the region defined by the mask, i.e., either the
union of temperature and polarization common masks for stacking on temperature
extrema or simply the polarization common masks when stacking on $E$-mode
extrema. These masks remove the Galactic plane, as well as the brighter point sources.
We define a 5$\deg\times$5$\deg$ grid, with the size of each pixel being
$0\fdg1$ and the number of pixels being 2500.  When adding $Q$ and $U$ images,
we weight each pixel by the number of unmasked $N_{\rm side} = 1024$ pixels
that lie in each re-gridded pixel (which is not uniform, because of the
re-gridding and masking; this weighting is used in the estimation of the
covariance matrix of the stacked images).  Therefore the pixels near the centre
generally have somewhat lower noise in the final stacked image. We then
generate $Q^{T,E}_{\rm r}$ and $U^{T,E}_{\rm r}$ images using Eqs.~\eqref{eq:def_qr}
and~\eqref{eq:def_ur}.

The predictions for $Q_r$ and $U_r$ are found by combining
Eqs.~\eqref{eq:qrprofile}--\eqref{eq:Eurprofile} with
Eqs.~\eqref{eq:ee}--\eqref{eq:eb}:
\begin{align}
  \left< Q^T_{\rm r} \right> (\theta) = - \cos{(2\alpha)} &\int \frac{\ell d\ell}{2\pi}
  W^T_\ell W^P_\ell \notag\\
  &\left(\bar{b}_\nu + \bar{b}_\zeta \ell^2\right) C^{TE}_\ell J_2(\ell\theta);
  \label{eq:qrmodel}\\
  \left< U^T_{\rm r} \right> (\theta) = - \sin{(2\alpha)} &\int \frac{\ell d\ell}{2\pi}
  W^T_\ell W^P_\ell \notag\\
  &\left(\bar{b}_\nu + \bar{b}_\zeta \ell^2\right) C^{TE}_\ell J_2(\ell\theta).
  \label{eq:urmodel}
\end{align}
For stacking on $E$-modes we have
\begin{align}
 \left\langle Q^E_{\rm r} \right\rangle (\theta) =
  -& \int \frac{\ell d\ell}{2\pi} W^E_{\ell} W^P_{\ell}
  \left(\bar{b}_\nu + \bar{b}_\zeta \ell^2\right) \notag\\
  & \left(C^{EE}_{\ell} \cos^2{(2\alpha)} +
  N^{EE}_{\ell}\right) J_2(\ell\theta),
  \label{eq:Eqrmodel}\\
 \left\langle U^E_{\rm r} \right\rangle (\theta) = -& \frac{1}{2} \sin{(4\alpha)}
  \int \frac{\ell d\ell}{2\pi} W^E_{\ell} W^P_{\ell} \notag \\
  & \left(\bar{b}_\nu + \bar{b}_\zeta {\ell}^2\right)
  C^{EE}_{\ell} J_2({\ell}\theta), \label{eq:Eurmodel}
\end{align}
where we have assumed that $C^{BB}_{\ell} = 0$, which is consistent with
our data, since \Planck\ does not have a direct detection of $B$-modes.  We use
a uniform prior on $\alpha$, $P(\alpha)$, when sampling the likelihood,
i.e.,
\begin{align}
  P(\alpha | d) &\propto P(d | \alpha) P(\alpha),
  \label{eq:bayes}
\end{align}
with
\begin{align}
  P(d|\alpha) &= \frac{1}{\sqrt{2\pi|\tens{C}|}} e^{-\frac{1}{2} \{d - (Q_{\rm
  r},U_{\rm r})(\alpha)\}^{\textsf T} \tens{C}^{-1} \{d - (Q_{\rm r}, U_{\rm
  r})(\alpha)\}}.
  \label{eq:probdata}
\end{align}
Here $d$ represents the data, consisting of the stacked $Q_{\rm r}$ and $U_{\rm
r}$ images, and $(Q_{\rm r}, U_{\rm r})(\alpha)$ are the predictions as a
function of $\alpha$ (see Eqs.~\ref{eq:qrmodel}--\ref{eq:Eurmodel}). The
quantity $\tens{C}$ is the covariance matrix, which is a combination of the
noise in the data and the cosmic variance due to the limited number of hot
(or cold) spots in the sky.  We have estimated the covariance matrix by
determining an rms level from the pixelization scheme chosen and then weighting this
with the inverse of the total number of pixels used in each re-gridded pixel;
we have also assumed that the covariance is diagonal in pixel space.

For the purposes of evaluating the likelihood, we have fixed $C^{TE}_{\ell}$
and $C^{EE}_{\ell}$ to the theoretical power spectra, based on the best-fit
\Planck\ parameters \citep{planck2014-a15}, and simply evaluate the likelihood
in a fine grid of $\alpha$ values. The choice of fixing the angular power
spectra is reasonable because the usual cosmological parameters are determined by
$C^{TT}_{\ell}, C^{TE}_{\ell},$ and $C^{EE}_{\ell}$, which are minimally
affected by $\alpha$ (no dependence, quadratic, and still quadratic dependence on
$\alpha$, respectively). See also comments at the end of Sect. \ref{sec_dataset_and_simulations}.

Finally, we quote the mean of the posterior on $\alpha$ and the width of the
posterior containing 68\,\% of the likelihood as the best-fit and statistical
uncertainty, respectively. The posterior for $\alpha$ is sufficiently Gaussian
that these two values contain all necessary information about the posterior.

\begin{figure}[h]
\begin{center}
\mbox{\includegraphics[width=\hsize]{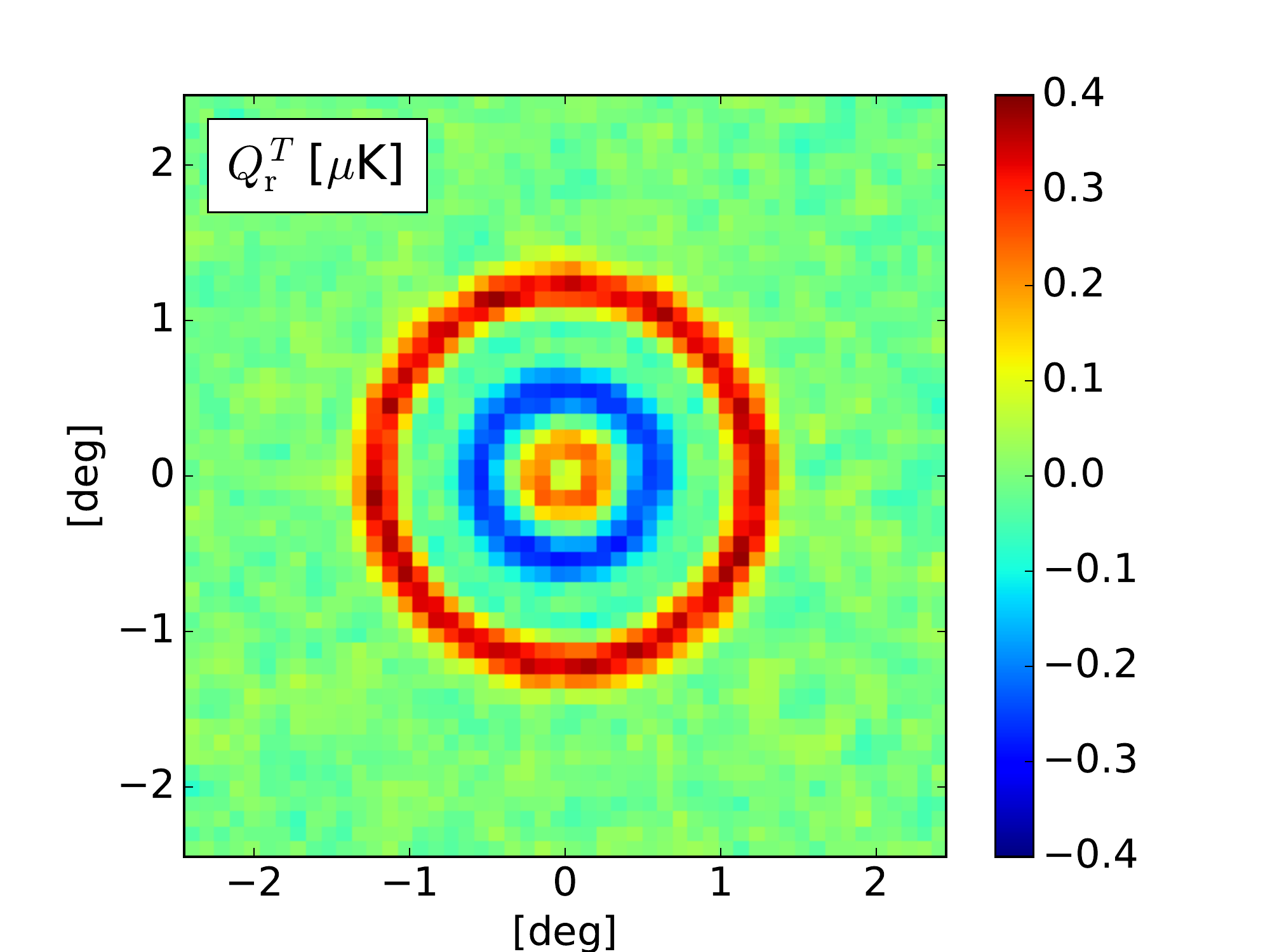}}
\mbox{\includegraphics[width=\hsize]{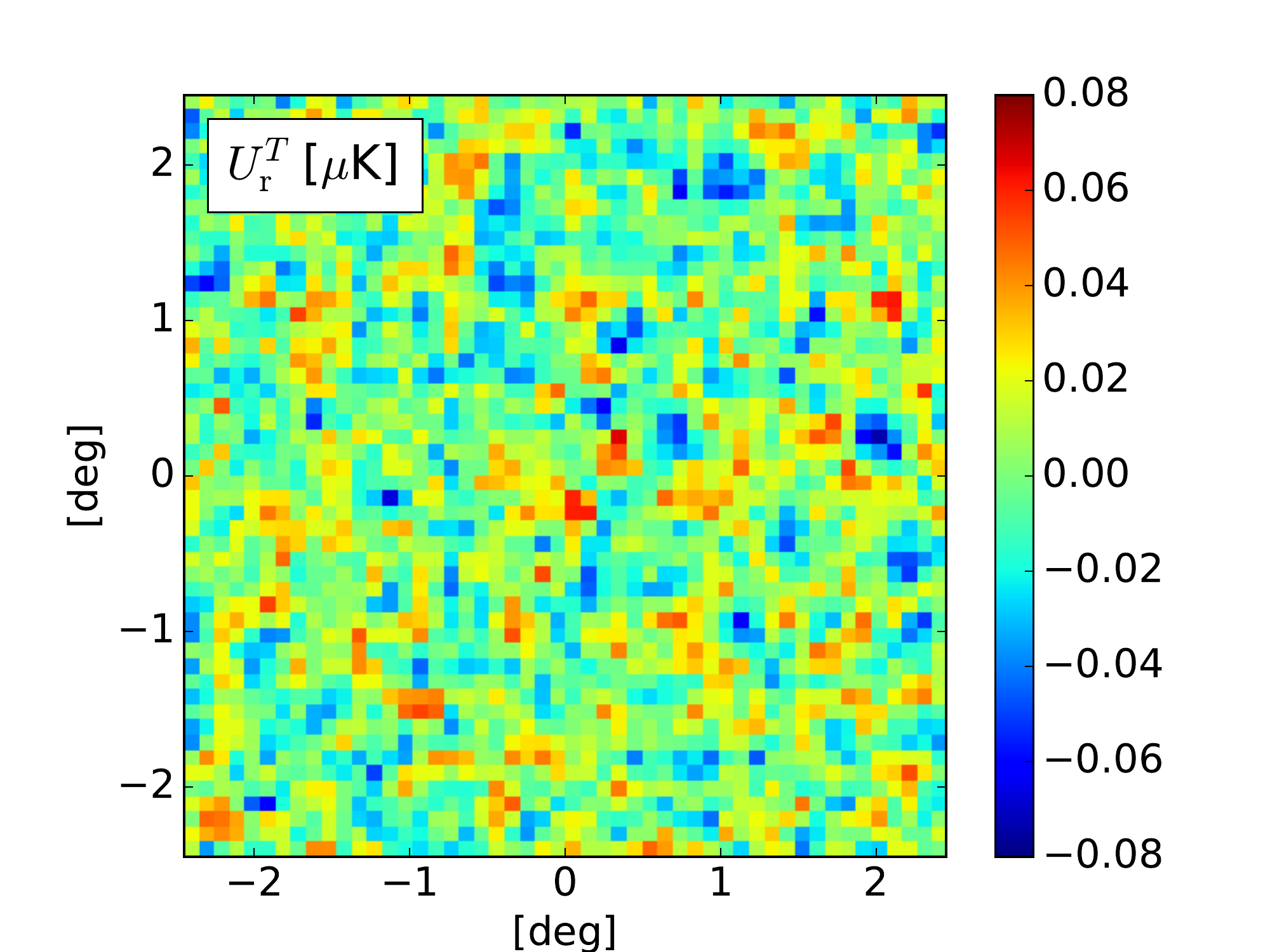}}
\end{center}
\caption{Stacked images of the transformed Stokes parameters $Q_{\rm r}$ (top) and
$U_{\rm r}$ (bottom) for \commander\ temperature hot spots. The rotation of the plane
of polarization will act to leak the signal from $Q_{\rm r}$ into $U_{\rm r}$. Note that
the bottom plot uses a different colour scale to enhance any weak features.
Finer resolution stacked images can be seen in figure~40 of
\citet{planck2014-a18}.}
\label{fig:Tstacks}
\end{figure}

\begin{figure}[h]
\begin{center}
\mbox{\includegraphics[width=\hsize]{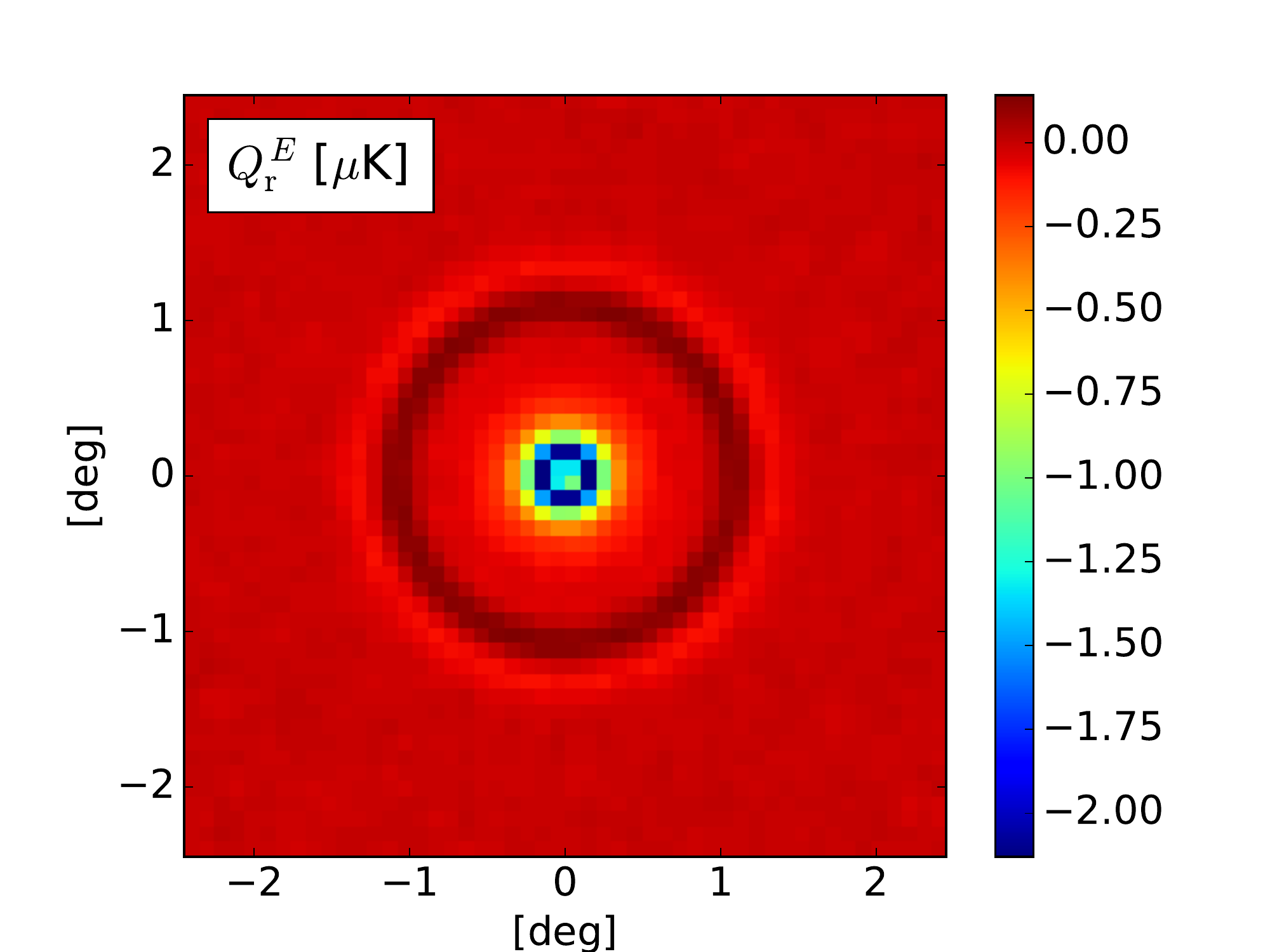}}
\mbox{\includegraphics[width=\hsize]{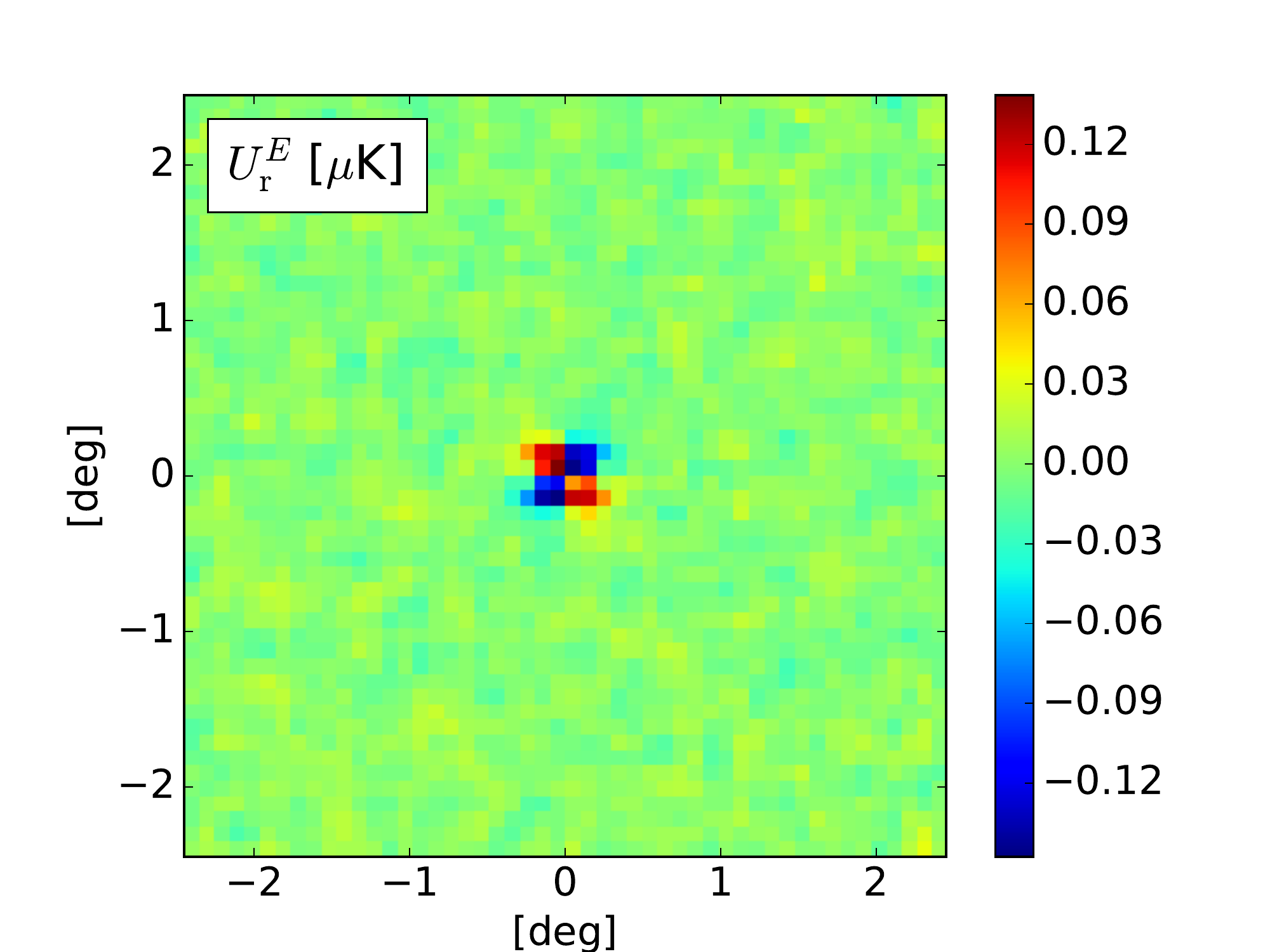}}
\end{center}
\caption{Stacked images of the transformed Stokes parameters $Q_{\rm r}$ (top)
and $U_{\rm r}$ (bottom) for \smica\ $E$-mode hot spots. The rotation of the
plane of polarization will act to leak the signal from $Q_{\rm r}$ into $U_{\rm
r}$. The quadrupole pattern in the bottom plot is related to ``sub-pixel''
effects \citep{planck2013-p08, planck2014-a13}; fortunately, our results are
insensitive to this feature, because it disappears in an azimuthal average (see
Sect.~\ref{sec:stackresults}).}
\label{fig:Estacks}
\end{figure}

\subsection{Harmonic-space analysis}

The harmonic-based analysis uses the so-called $D$-estimators \citep[see for
instance][]{Wu2009,Gruppuso:2011ci,Zhao:2015mqa,Gruppuso:2016nhj}, which are defined by the
following equations:
\begin{eqnarray}
  D_{\ell}^{TB, {\rm obs}} &=& C'^{TB}_{\ell} \cos (2 \hat \alpha) - C'^{TE}_{\ell} \sin (2  \hat \alpha)  \, ; \label{DTB} \\
  D_{\ell}^{EB, {\rm obs}} &=& C'^{EB}_{\ell} \cos(4  \hat \alpha) - \frac{1}{2} (C'^{EE}_{\ell} - C'^{BB}_{\ell}) \sin(4  \hat \alpha) \, . \label{DEB}
\end{eqnarray}
Here $ \hat \alpha$ is the estimate for the birefringence angle $\alpha$. It is possible to show that on average
\begin{eqnarray}
\langle D_{\ell}^{TB, {\rm obs}} \rangle &=& \langle C_{\ell}^{TE} \rangle \sin (2 (\alpha -  \hat \alpha)) \, , \label{aveTBbeta} \\
\langle D_{\ell}^{EB, {\rm obs}} \rangle &=& \frac{1}{2} \, \left( \langle C_{\ell}^{EE} \rangle - \langle C_{\ell}^{BB} \rangle \right) \, \sin (4 (\alpha -  \hat \alpha)) \, . \label{aveEBbeta}
\end{eqnarray}
Eqs.~(\ref{aveTBbeta}) and (\ref{aveEBbeta}) are zero when
\begin{equation}
 \hat \alpha = \alpha \, .
\label{betaugualealpha}
\end{equation}
Eq.~(\ref{betaugualealpha}) suggests that we can find $\alpha$ by looking for
the $\hat \alpha$ that
makes null the expectation values of the $D$-estimators.
From now on we always consider that Eq.~(\ref{betaugualealpha}) is satisfied.

We estimate the angular power spectra using the {\tt MASTER} method \citep{Hivon2002}
extended to polarization \citep{kogut2003,Polenta2005} to correct for masking,
and we use simulations to estimate the noise. We choose a bin size of
$\Delta\ell = 20$, starting at $\ell_{\min} = 51$, to avoid correlations between
bins induced by masking.  It is then possible to minimize $\chi^2(\alpha)$ for
$TB$ and $EB$ separately, or jointly to estimate $\alpha$:
\begin{equation}
\chi^2_X(\alpha) 
=  \sum_{b b^{\prime}} D^{X, {\rm obs}}_{b}  {M^{X X}_{b b^{\prime}}}^{-1}
D^{X, {\rm obs}}_{b^{\prime}}
\, ,
\label{loglike}
\end{equation}
where $X=TB$ or $EB$, $b$ denotes the bin
and
$M^{X X}_{b b^{\prime}} = 
\langle D^{X}_{b}  D^{X}_{b^{\prime}} \rangle$, where the average is taken over
the FFP8.1 simulations described in Sect.~\ref{sec_dataset_and_simulations} and
generated with $\alpha = 0$. We thus are adopting a simple frequentist approach
to test the null hypothesis of no parity violation.
This approach also allows for the minimization of Eq.~(\ref{loglike}) in
subintervals of multipoles, providing the possibility of
searching for a possible angular scale dependence
to the birefringence effect, i.e.,
\begin{equation}
\chi^2_X(\alpha)  =  \sum_{b} \chi^2_{X, b} (\alpha)
\, ,
\label{loglike1}
\end{equation}
where $ \chi^2_{X, b} (\alpha) = \sum_{b^{\prime}} D^{X, {\rm obs}}_{b}
{M^{X X}_{b b^{\prime}}}^{-1} D^{X, {\rm obs}}_{b^{\prime}} $.
This will be used to test the stability of the estimates of $\alpha$
against the ranges of multipoles considered for the CMB spectra.

The $D^{TB}$-estimator is inherently built from cross-correlations (see
Eq.~\ref{DTB}) and therefore we are able to use the full-mission data and
simulations for all component-separation methods to generate the corresponding $\chi^2$.
Moreover, since the \smica\ simulations are also delivered in half-mission
form, we are
additionally able to estimate $D^{TB}_{\ell}$ by cross-correlating half-mission 1 and
half-mission 2 data and simulations.

Regarding the $D^{EB}$-estimator, since it contains auto-correlations
(see Eq.~\ref{DEB}), we must estimate it from the half-mission data along with
simulations in order to satisfy
the recommendations on the use of polarization data given in
\citet{planck2014-a11}, based on the fact that only results with a weak
dependence on noise are to be considered fully reliable.
More specifically, $C'^{EE}_{\ell}$ and $C'^{BB}_{\ell}$ are estimated from cross-correlating half-mission 1 with
half-mission 2 \smica\ data and using the corresponding \smica\ simulations only.

\section{Results}
\label{sec_results}

In the following subsections we present our constraints on $\alpha$ for the two
methods, described in the previous section. We will quote our best-fit $\alpha$
values and uncertainties (statistical only, leaving consideration of systematic
effects to
Sect.~\ref{sec:systematics}). We will show specifically that the $E$--$B$
correlation is more constraining than the $T$--$B$ correlation. This is
expected and can be demonstrated directly by computing the variance of
Eqs.~\eqref{eq:tb} and \eqref{eq:eb}. For small $\alpha$, the variance of
$\alpha$ based on $T$--$B$ and $E$--$B$ information alone is
\begin{align}
  (2\ell + 1)f_{\rm sky} (\sigma^{TB}_{\ell})^2 &\simeq \frac{1}{4}
  \frac{C^{TT}_{\ell}C^{BB}_{\ell}}{(C^{TE}_{\ell})^2} \gtrsim \frac{1}{4}
  \frac{C^{BB}_{\ell}}{C^{EE}_{\ell}},
  \label{eq:tbuncertainty}\\
  (2\ell + 1)f_{\rm sky} (\sigma^{EB}_{\ell})^2 &\simeq \frac{1}{4}
  \frac{C^{EE}_{\ell}C^{BB}_{\ell}}{(C^{EE}_{\ell} - C^{BB}_{\ell})^2}
  \simeq \frac{1}{4} \frac{C^{BB}_{\ell}}{C^{EE}_{\ell}},
  \label{eq:ebuncertainty}
\end{align}
respectively. This can be derived from the Fisher information matrix, where the
covariance is a simple $1\times1$ matrix containing the variance of $T$--$B$ or
$E$--$B$. Thus, as suggested by the above relations, our results based on
$E$--$B$ are generally more constraining than our $T$--$B$ results (the
presence of noise, however, will modify these relations).

\begin{figure}
\centering
\includegraphics[width=9cm]{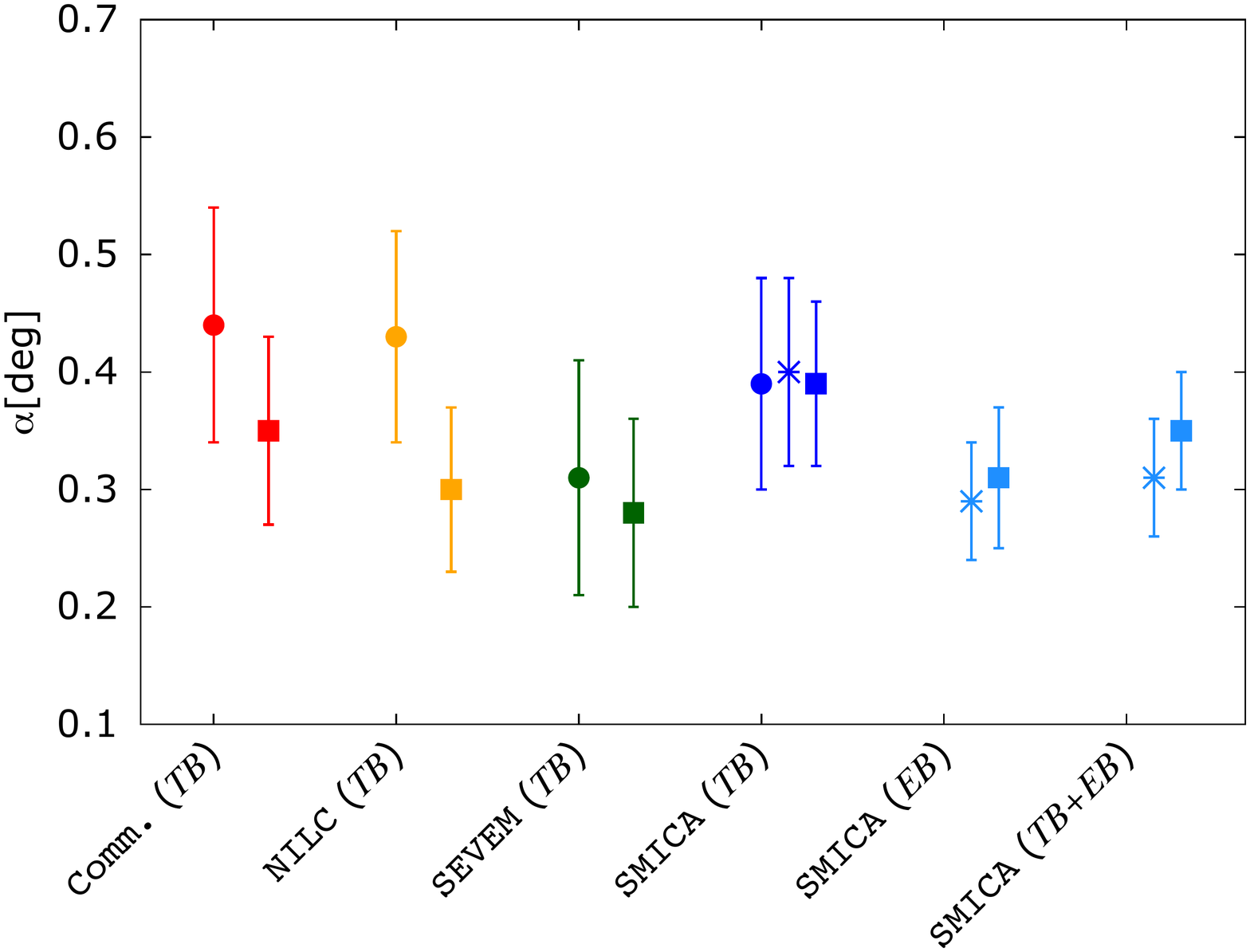}
\caption{Comparison between harmonic and pixel-based analysis.
Dot (star) symbols show the estimates coming from $D$-estimators built with CMB spectra obtained from
full-mission (half-mission) data.
Square symbols represent the estimates coming from the stacking analysis.}
\label{three}
\end{figure}

With respect to statistical uncertainty, we will demonstrate that our results
are robust to
all component-separation methods,
and with respect to our two methods.  For
convenience, we offer a direct comparison of results obtained by our two
approaches in Fig.~\ref{three}.

\subsection{Map-space results}
\label{sec:stackresults}

Firstly we note that as a basic check we have verified that
Fig.~\ref{fig:Tstacks} closely reproduces the stacked images shown in
\citet{planck2014-a18}. We also show the $Q_{\rm r}$ and $U_{\rm r}$ images stacked on
$E$-mode extrema in Fig.~\ref{fig:Estacks}. The visually striking quadrupole
pattern in $U^{E}_{\rm r}$ appears to be an artefact of the
pixelization scheme and is related to the so-called ``sub-pixel'' effects
described in \citet{planck2013-p08} and \citet{planck2014-a13}. This happens because
the pixels of the stacked $Q$ image are imperfectly separated near
the centre of the map (the stacked $U$ image does not exhibit this imperfect
mixing, because the pixel boundaries align perfectly with where the profile
changes sign). The pixelization errors are more evident in the $U^E_{\rm r}$ image
than the $U^T_{\rm r}$ because the individual $Q^E$ and $U^E$ images are strongly
peaked near the centre of the image, and thus when generating the $U_{\rm r}$
stack
imperfect subtraction leads to features in the centre of Fig.~\ref{fig:Estacks}
(bottom). This effect has a non-diagonal influence on the power spectra and
thus has a negligible effect on parameters \citep{planck2013-p08} and this
analysis. Alternatively, since constraints on $\alpha$ come only from the
radial part of the stacked images, the pixelization pattern seen in the centre
of Fig.~\ref{fig:Estacks}, which cancels out in the azimuthal average, will not
bias our $\alpha$ results (though it will contribute to the statistical
uncertainty).

\begin{figure}[h]
\begin{center}
\mbox{\includegraphics[width=\hsize]{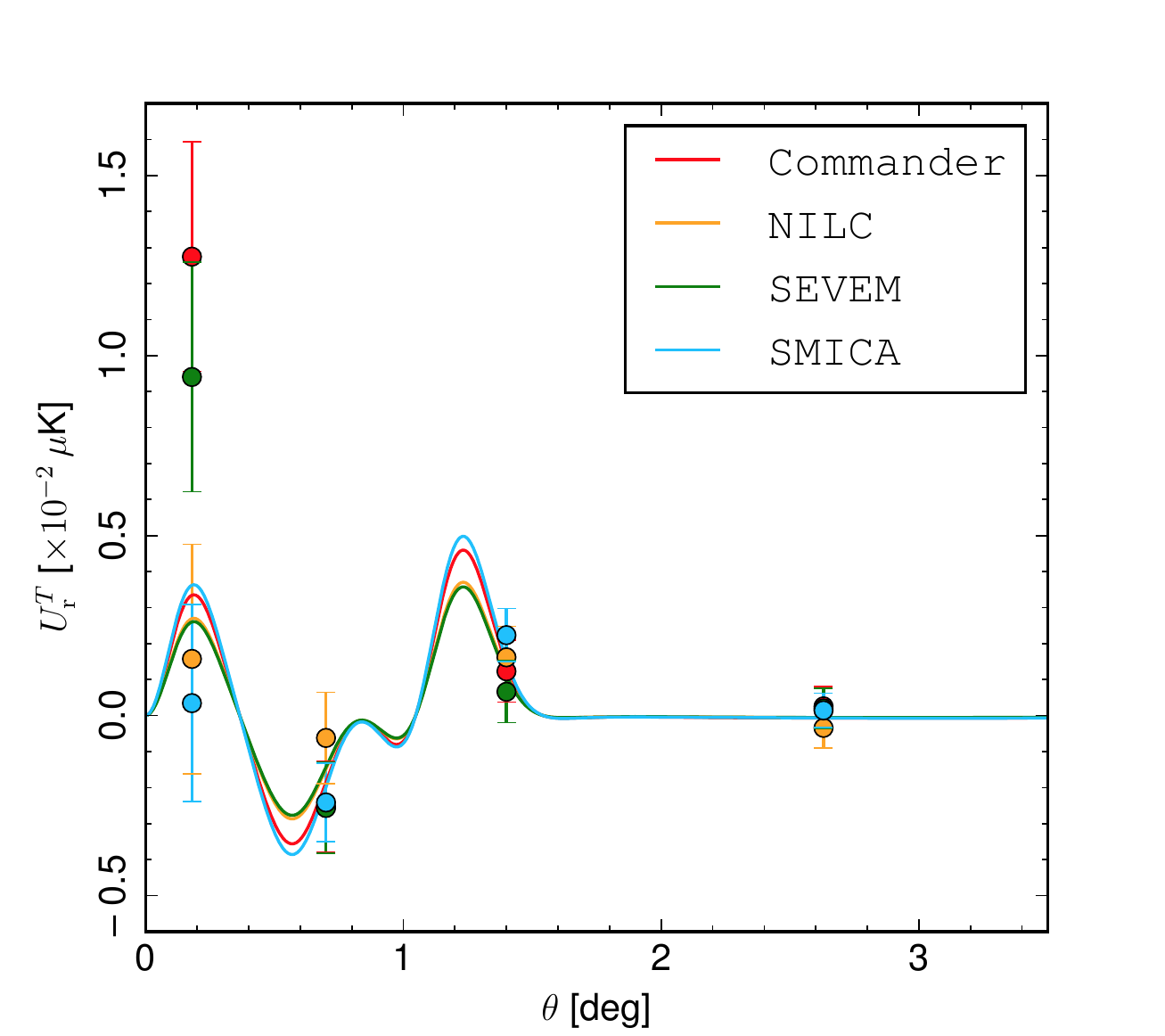}}
\mbox{\includegraphics[width=\hsize]{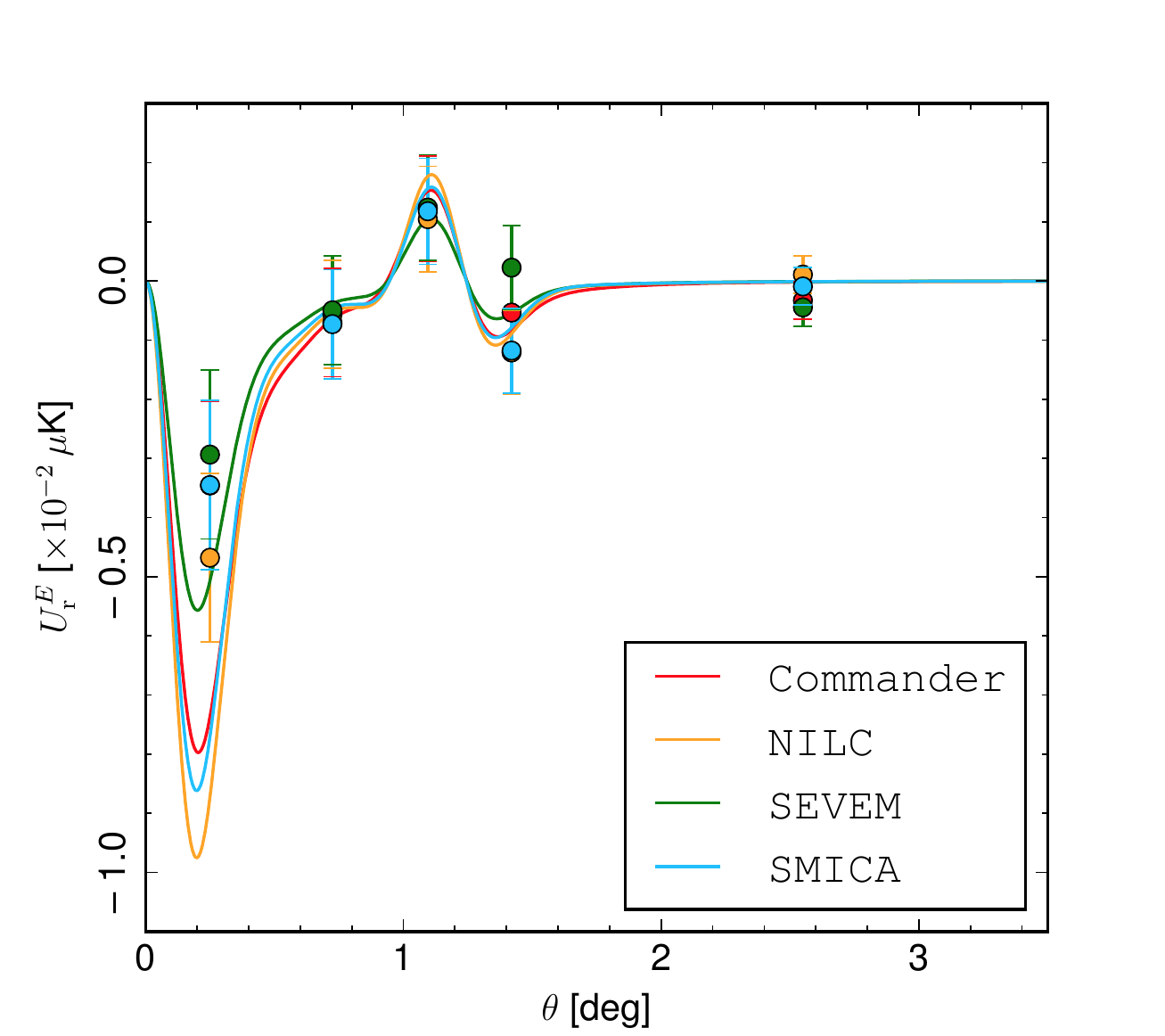}}
\end{center}
\caption{Profiles of $U_{\rm r}$ from stacking on temperature (top) and $E$-mode
(bottom) extrema for the four component-separation methods. The best-fit curves
for each component-separation method are also
shown, with $\alpha$ values given in the fourth
column of Table~\ref{tab:constraints}. Note that we have included both hot and
cold spots in this figure, i.e., we have co-added the negative of the profile
from cold spots to the profile of the hot spots. Error bars correspond to
68\,\% confidence regions.}
\label{fig:profile}
\end{figure}

Fig.~\ref{fig:profile} shows the binned $U_{\rm r}$ profiles for the four
component-separation methods.
The apparently non-zero $\alpha$ signal seen in Fig.~\ref{fig:profile}
is not visible in Figs.~\ref{fig:Tstacks} and \ref{fig:Estacks}. This is
mainly due to the fact that any signal in the stacked images must be
shared out over the 2500 pixels and partially due to the fact that $U_{\rm r}$
oscillates about zero for $\alpha \neq 0$.
The binning here is chosen to pick out ranges
with the same sign in the predicted curve for $\alpha \neq 0\deg$ (with the
$\Lambda$CDM prediction being identically zero); this choice is for
visualization purposes only, since the statistical fit is performed on
the original stacked images, i.e., Figs.~\ref{fig:Tstacks} and
\ref{fig:Estacks}.

Results are summarized in Table~\ref{tab:constraints} and \ref{tab:chisquareds}
for \commander, \nilc, \sevem, and \smica. Table~\ref{tab:constraints} contains
the constraint on $\alpha$ based on the high-pass-filtered $Q$ and $U$ maps and
their half-mission half differences (HMHD), which give a useful measure of the
noise in the data. We present results based on stacking on temperature and
$E$-mode extrema, both separately and combined. We have estimated that the
correlation of the temperature and $E$-mode stacks are at the sub-percent level
by looking at the amount of overlap in the positions of the peaks; thus we can
safely neglect correlations in the combined fit.
From Table~\ref{tab:chisquareds} we can see that for most cases
$\alpha=0$ fits the data reasonably well; however, the reduction in $\chi^2$
from a non-zero $\alpha$ is large enough, compared to the expectation of adding
a single parameter, to yield a significant detection (with respect to
statistical uncertainty only). In other words, while a horizontal line going through
$0\,\mu$K might seem like an acceptable fit in
Fig.~\ref{fig:profile},
a non-zero $\alpha$ is able to pick out the oscillatory features providing a
significantly better fit.
We report $5$--$7\,\sigma$
detections for $\alpha$ (with respect to statistical uncertainty only),
however, this can be completely explained by a systematic rotation of our
polarization-sensitive bolometers (PSBs) which we discuss in
Sect.~\ref{sec:systematics}. Null-test estimates all give $\alpha$ within
1$\,\sigma$ of 0$\deg$, with the exception of \commander\ results stacked on
temperature, which are slightly above $1\,\sigma$ (see the second and
eighteenth rows of Table~\ref{tab:constraints}). We have also checked that
there is very weak dependence on our results coming from the different choices
for the thresholds used to define the peaks.

\begin{table}
\begingroup
\newdimen\tblskip \tblskip=5pt
\caption{Mean values and ($1\,\sigma$) statistical uncertainties for $\alpha$ (in degrees)
derived from the stacking analysis for all component-separation methods, coming
from hot spots, cold spots, and all extrema.}
\label{tab:constraints}
\nointerlineskip
\vskip -3mm
\footnotesize
\setbox\tablebox=\vbox{
   \newdimen\digitwidth
   \setbox0=\hbox{\rm 0}
   \digitwidth=\wd0
   \catcode`*=\active
   \def*{\kern\digitwidth}
   \newdimen\signwidth
   \setbox0=\hbox{+}
   \signwidth=\wd0
   \catcode`!=\active
   \def!{\kern\signwidth}
\halign{\tabskip 0pt \hbox to 1.0in{#\leaderfil}\tabskip 4pt&
         \hfil#\hfil\tabskip 8pt&
         \hfil#\hfil\tabskip 8pt&
         \hfil#\hfil\tabskip 0pt\cr                           
\noalign{\doubleline\vskip 1pt}
\omit\hfil Method\hfil& !Hot& !Cold& !All\cr   
\noalign{\vskip 4pt\hrule\vskip 6pt}
\multispan4 \hfil $T$--$B$ \hfil\cr
\noalign{\vskip 3pt}
{\tt \commander}& $!0.36\pm0.12$& $!0.34\pm0.11$& $!0.35\pm0.08$\cr
\noalign{\vskip 3pt}
    HMHD$^{\rm a}$& $-0.13\pm0.12$& $-0.20\pm0.11$& $-0.16\pm0.08$\cr
\noalign{\vskip 3pt}
     {\tt \nilc}$^{\rm b}$& $!0.23\pm0.10$& $!0.36\pm0.10$& $!0.30\pm0.07$\cr
\noalign{\vskip 3pt}
    HMHD$^{\rm a}$& $-0.08\pm0.10$& $!0.02\pm0.10$& $-0.03\pm0.07$\cr
\noalign{\vskip 3pt}
    {\tt \sevem}& $!0.37\pm0.12$& $!0.18\pm0.12$& $!0.28\pm0.08$\cr
\noalign{\vskip 3pt}
    HMHD$^{\rm a}$& $!0.07\pm0.12$& $!0.07\pm0.12$& $!0.07\pm0.08$\cr
\noalign{\vskip 3pt}
    {\tt \smica}$^{\rm b}$& $!0.42\pm0.10$& $!0.36\pm0.10$& $!0.39\pm0.07$\cr
\noalign{\vskip 3pt}
    HMHD$^{\rm a}$& $-0.04\pm0.10$& $-0.04\pm0.10$& $-0.04\pm0.07$\cr
\noalign{\vskip 3pt}
\multispan4 \hfil $E$--$B$ \hfil\cr
\noalign{\vskip 3pt}
 {\tt \commander}& $!0.41\pm0.11$& $!0.44\pm0.11$& $!0.43\pm0.08$\cr
\noalign{\vskip 3pt}
     HMHD$^{\rm a}$& $!0.03\pm0.11$& $-0.07\pm0.11$& $-0.02\pm0.08$\cr
\noalign{\vskip 3pt}
      {\tt \nilc}$^{\rm b}$& $!0.33\pm0.09$& $!0.38\pm0.08$& $!0.35\pm0.06$\cr
\noalign{\vskip 3pt}
     HMHD$^{\rm a}$& $-0.10\pm0.09$& $!0.01\pm0.08$& $-0.05\pm0.06$\cr
\noalign{\vskip 3pt}
     {\tt \sevem}& $!0.28\pm0.12$& $!0.32\pm0.12$& $!0.30\pm0.09$\cr
\noalign{\vskip 3pt}
     HMHD$^{\rm a}$& $!0.04\pm0.12$& $!0.04\pm0.12$& $!0.04\pm0.09$\cr
\noalign{\vskip 3pt}
     {\tt \smica}$^{\rm b}$& $!0.25\pm0.09$& $!0.37\pm0.09$& $!0.31\pm0.06$\cr
\noalign{\vskip 3pt}
     HMHD$^{\rm a}$& $-0.11\pm0.09$& $!0.01\pm0.09$& $-0.05\pm0.06$\cr
\noalign{\vskip 3pt}
\multispan4 \hfil Combined \hfil\cr
\noalign{\vskip 3pt}
 {\tt \commander}& $!0.38\pm0.08$& $!0.40\pm0.08$& $!0.39\pm0.06$\cr
\noalign{\vskip 3pt}
     HMHD$^{\rm a}$& $-0.05\pm0.08$& $-0.12\pm0.08$& $-0.09\pm0.06$\cr
\noalign{\vskip 3pt}
      {\tt \nilc}$^{\rm b}$& $!0.28\pm0.06$& $!0.37\pm0.06$& $!0.33\pm0.05$\cr
\noalign{\vskip 3pt}
     HMHD$^{\rm a}$& $-0.10\pm0.06$& $!0.01\pm0.06$& $-0.04\pm0.05$\cr
\noalign{\vskip 3pt}
     {\tt \sevem}& $!0.32\pm0.08$& $!0.25\pm0.08$& $!0.29\pm0.06$\cr
\noalign{\vskip 3pt}
     HMHD$^{\rm a}$& $!0.05\pm0.09$& $!0.05\pm0.08$& $!0.05\pm0.06$\cr
\noalign{\vskip 3pt}
     {\tt \smica}$^{\rm b}$& $!0.32\pm0.07$& $!0.37\pm0.06$& $!0.35\pm0.05$\cr
\noalign{\vskip 3pt}
     HMHD$^{\rm a}$& $-0.08\pm0.07$& $-0.01\pm0.06$& $-0.04\pm0.05$\cr
\noalign{\vskip 3pt\hrule\vskip 4pt}}}
\endPlancktable                    
\tablenote {{\rm a}} We include the fit from each component-separation method's
half-mission half-difference (HMHD) $Q$ and $U$ maps, as an indication of the
expectation for noise.\par
\tablenote {{\rm b}} \nilc\ and \smica\ have smaller uncertainties compared with
\commander\ and \sevem, which follows from the naive expectation of the rms in
the polarization maps \citep[see table~1 of][]{planck2014-a11}.\par
\endgroup
\end{table}

\begin{table*}
\begingroup
\newdimen\tblskip \tblskip=5pt
\caption{$\chi^2$ values for the model with $\alpha = 0$, derived from the
stacking analysis for all component-separation methods. The $\Delta\chi^2$ is
the reduction of $\chi^2$ given the values of $\alpha$ in the corresponding
entry in table~\ref{tab:constraints}. For convenience we have also included
the probability to exceed (PTE) for each value of $\alpha$.}
\label{tab:chisquareds}
\nointerlineskip
\vskip -3mm
\footnotesize
\setbox\tablebox=\vbox{
   \newdimen\digitwidth
   \setbox0=\hbox{\rm 0}
   \digitwidth=\wd0
   \catcode`*=\active
   \def*{\kern\digitwidth}
   \newdimen\signwidth
   \setbox0=\hbox{+}
   \signwidth=\wd0
   \catcode`!=\active
   \def!{\kern\signwidth}
\halign{\tabskip 0pt \hbox to 1.0in{#\leaderfil}\tabskip 4pt&
         \hfil#\hfil\tabskip 8pt&
         \hfil#\hfil\tabskip 8pt&
         \hfil#\hfil\tabskip 8pt&
         \hfil#\hfil\tabskip 8pt&
         \hfil#\hfil\tabskip 8pt&
         \hfil#\hfil\tabskip 0pt\cr                           
\noalign{\doubleline\vskip 1pt}
\omit&\multispan3 \hfil Hot \hfil & \multispan3 \hfil Cold \hfil \cr
\noalign{\vskip -5pt}
\omit&\multispan6\hrulefill\cr
\omit\hfil Method \hfil& $\chi^{2 \, \rm a}$ &  $\Delta\chi^2$& PTE & $\chi^{2 \, \rm a}$ &  $\Delta\chi^2$& PTE \cr   
\noalign{\vskip 4pt\hrule\vskip 6pt}
\multispan7 \hfil $T$--$B$ \hfil\cr
\noalign{\vskip 3pt}
{\tt \commander}               & $2453.7$ & $-9.2$   & $1.2\times10^{-3}$ & $2769.8$ & $-9.3$   & $1.1\times10^{-3}$ \cr
\noalign{\vskip 3pt}
     {\tt \nilc}               & $2525.8$ & $-5.2$   & $1.1\times10^{-2}$ & $2641.3$ & $-13.3$  & $1.3\times10^{-4}$ \cr
\noalign{\vskip 3pt}
    {\tt \sevem}               & $2552.6$ & $-9.7$   & $9.4\times10^{-4}$ & $2718.9$ & $-2.3$   & $6.4\times10^{-2}$ \cr
\noalign{\vskip 3pt}
    {\tt \smica}               & $2567.7$ & $ -17.2$ & $1.7\times10^{-5}$ & $2610.1$ & $ -13.0$ & $1.5\times10^{-4}$ \cr
\noalign{\vskip 3pt}
\multispan7 \hfil $E$--$B$ \hfil\cr
\noalign{\vskip 3pt}
 {\tt \commander}              & $2548.8$ & $-12.1$  & $2.5\times10^{-4}$ & $2542.7$ & $-15.9$  & $3.3\times10^{-5}$ \cr
\noalign{\vskip 3pt}
      {\tt \nilc}              & $2554.4$ & $-13.3$  & $1.3\times10^{-4}$ & $2555.9$ & $-19.3$  & $5.7\times10^{-6}$ \cr
\noalign{\vskip 3pt}
     {\tt \sevem}              & $2551.5$ & $-4.6$   & $1.6\times10^{-2}$ & $2552.0$ & $-6.7$   & $4.8\times10^{-3}$ \cr
\noalign{\vskip 3pt}
     {\tt \smica}              & $2556.8$ & $-7.9$   & $2.5\times10^{-3}$ & $2559.1$ & $-17.2$  & $1.7\times10^{-5}$ \cr
\noalign{\vskip 3pt\hrule\vskip 4pt}}}
\endPlancktablewide                 
\tablenote {{\rm a}} The number of degrees of freedom is 2500 coming from a $5\deg\times5\deg$ patch
with $0\fdg1$ pixel size.\par
\endgroup
\end{table*}

A stacking analysis similar to ours has been attempted in
\citet{Contaldi2015} also using \Planck\ data. Our results based on $E$--$B$
data are consistent with those of \citet{Contaldi2015}, but with smaller
statistical uncertainties; however, we disagree with \citet{Contaldi2015}
regarding the constraints coming from $T$--$B$ data, which are claimed to be
too noisy to be used. We show here that both $T$--$B$ and $E$--$B$ data can be
successfully exploited to constrain the birefringence angle.

\subsection{Harmonic-space results}

\begin{figure}
\centering
\includegraphics[width=\hsize]{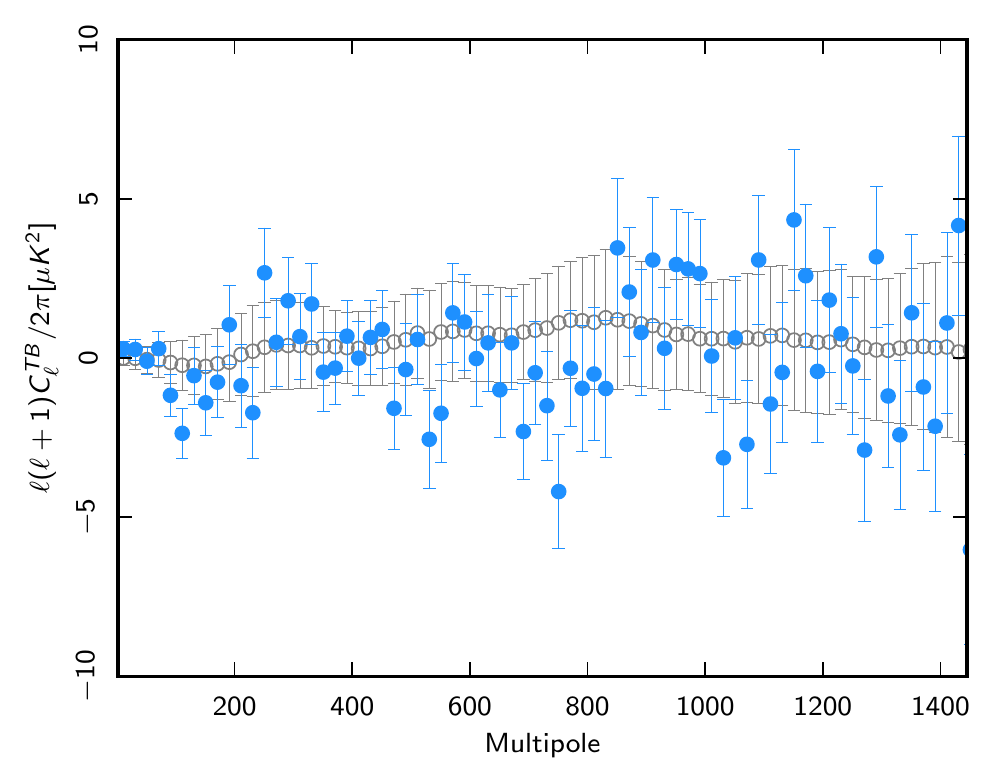}
\includegraphics[width=\hsize]{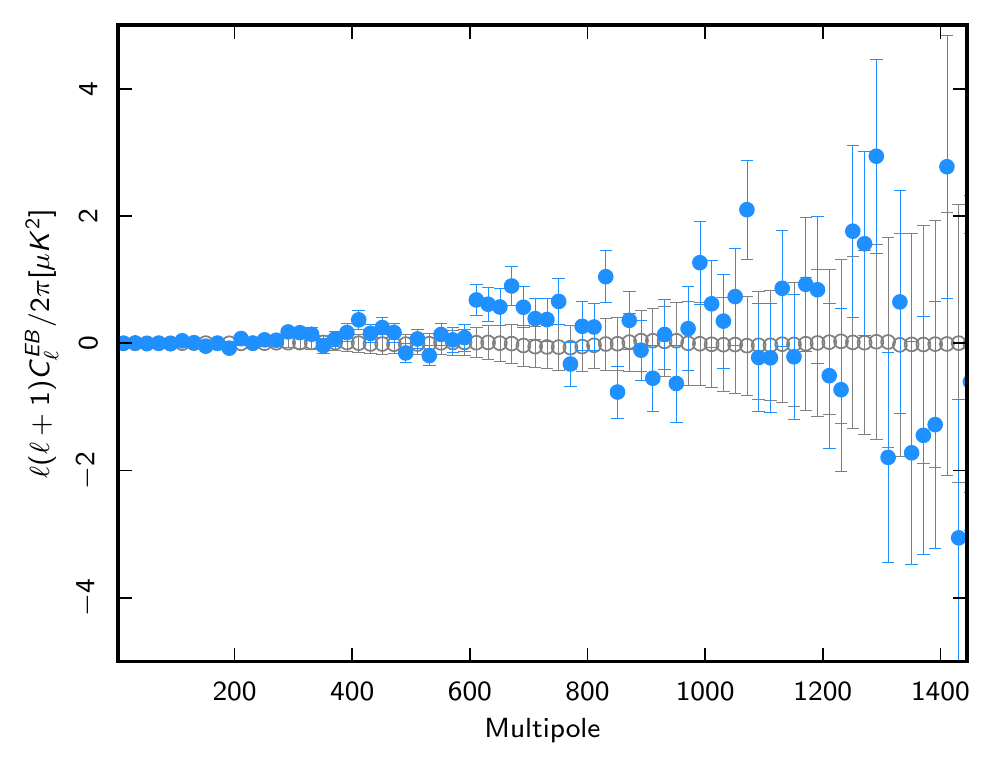}
\caption{Angular power spectrum estimates for $TB$ (top) and $EB$ (bottom),
with \smica\ data in blue and the corresponding simulations in black. Only
statistical uncertainties are shown here.
}
\label{smicaspectrahm}
\end{figure}

Following the recommendation given by the \Planck\ collaboration
\citep{planck2014-a11}, we present results below based only on cross-correlations.
Therefore, as described in
Sect.~\ref{sec_dataset_and_simulations}, we present results for the
$D^{TB}$-estimator using the full-mission data from \commander, \nilc, \sevem,
and \smica, and with half-mission data for \smica. For the same reasons we
present results with the $D^{EB}$-estimator using the half-mission data from
\smica\ only. Additionally we present a joint analysis with the half-mission data
from \smica. Of course the $D$-estimators are built through the CMB angular power spectra.
As an example, we display in Fig.~\ref{smicaspectrahm} the $TB$ and $EB$ CMB angular power spectra obtained
with the \smica\ method using half-mission data. We also show the FFP8.1 simulations
for the same component-separation method.

\subsubsection{$T$--$B$}

\begin{figure}
\centering
\includegraphics[width=9cm]{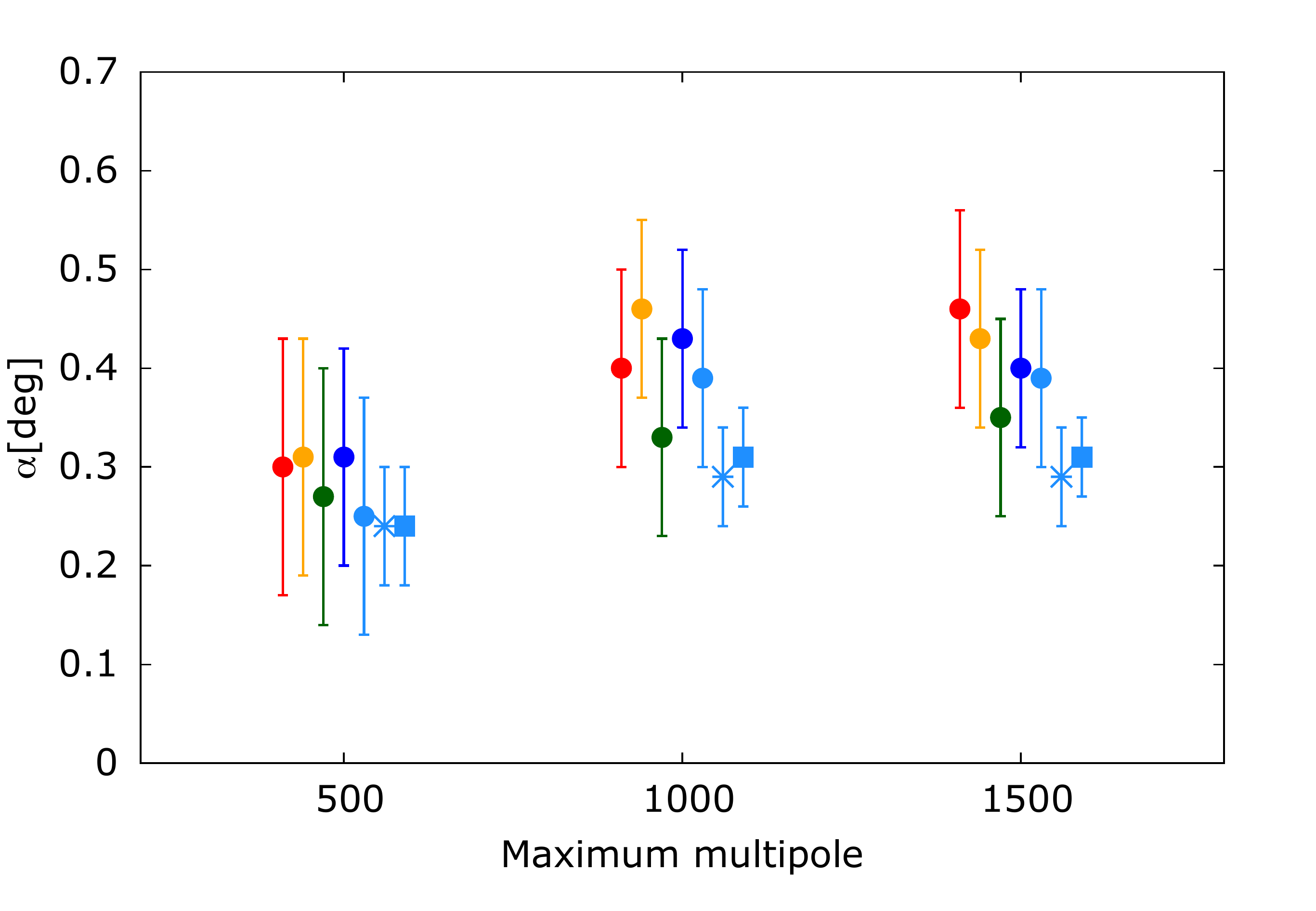}
\caption{Birefringence angle estimates (in degrees) versus the maximum multipole
considered. Only ($1\,\sigma$) statistical uncertainties are shown here; with
systematic errors discussed later.  \commander\ is shown in
red, \nilc\ in orange, \sevem\ in green, \smica\ (full-mission data) in blue,
and \smica\ (half-mission data) in cyan. Dot symbols refer to the estimates
obtained with the
$D_{\ell}^{TB}$ estimator. Star symbols refer to estimates coming from $D_{\ell}^{EB}$ and squares
are obtained through the combination of $D_{\ell}^{TB}$ and $D_{\ell}^{EB}$.}
\label{one}
\end{figure}

The estimates obtained are displayed in Fig.~\ref{one} as a function of the
maximum multipole considered ($\ell_{\max}$). We note that all the estimates
are stable among the component-separation methods and against the choice of
$\ell_{\max}$.  Moreover, the \smica\ results provide a further test of
stability with respect to computing the CMB angular power spectra from
full-mission or half-mission data and simulations.
The estimates of the birefringence angle for $\ell_{\max} \simeq 1500$ are also
reported in Table \ref{tableone} (see first five rows for $D_{\ell}^{TB}$).

\begin{table}
\begingroup
\newdimen\tblskip \tblskip=5pt
\caption{Minimum $\chi^2$ values and statistical uncertainties ($1\,\sigma$)
for $\alpha$,
derived from the $D$-estimators with $\ell_{\max} \simeq 1500$.
The $\chi^2$ values for $\alpha=0$ and the change $\Delta \chi^2$ for the corresponding
value of the birefringence angle are provided in the fourth and fifth columns
respectively.}
\label{tableone}
\nointerlineskip
\vskip -3mm
\footnotesize
\setbox\tablebox=\vbox{
   \newdimen\digitwidth
   \setbox0=\hbox{\rm 0}
   \digitwidth=\wd0
   \catcode`*=\active
   \def*{\kern\digitwidth}
   \newdimen\signwidth
   \setbox0=\hbox{+}
   \signwidth=\wd0
   \catcode`!=\active
   \def!{\kern\signwidth}
\halign{\tabskip 0pt \hbox to 1.0in{#\leaderfil}\tabskip 4pt&
         \hfil#\hfil\tabskip 4pt&
         \hfil#\hfil\tabskip 4pt&
         \hfil#\hfil\tabskip 4pt&
         \hfil#\hfil\tabskip 4pt\cr                           
\noalign{\doubleline\vskip 1pt}
\omit\hfil Method\hfil& !$\alpha\,$[deg]& !bias$^{\rm a}\,$[deg] & $\chi^2(\alpha=0)$ & $\Delta \chi^2$ $^{\rm d}\,$ \cr   
\noalign{\vskip 4pt\hrule\vskip 6pt}
\multispan5 \hfil $T$--$B$ \hfil\cr
\noalign{\vskip 3pt}
{\tt \commander}& $!0.44\pm0.10$& $!0.01$ & $87.1$ & $-20.9$!\cr
\noalign{\vskip 3pt}
     {\tt \nilc}$^{\rm b}$& $!0.43\pm0.09$& $-0.01$ & $104.3$ & $-22.5$!\cr
\noalign{\vskip 3pt}
    {\tt \sevem}& $!0.31\pm0.10$& $!0.02$ & $80.0$ & $-10.3!$\cr
\noalign{\vskip 3pt}
    {\tt \smica}$^{\rm b}$& $!0.40\pm0.08$& $!0.00$ & $92.7$ & $-23.9$!\cr
\noalign{\vskip 3pt}
    {\tt \smica}$\times^{\rm bc}$& $!0.39\pm0.09$& $-0.01$ & $92.8$ & $-18.8$!\cr
\noalign{\vskip 3pt}
\multispan5 \hfil $E$--$B$ \hfil\cr
\noalign{\vskip 3pt}
 {\tt \smica}$\times^{\rm bc}$& $!0.29\pm0.05$& $!0.00$ & $135.9$ & $-39.9$!\cr
\noalign{\vskip 3pt}
\multispan5 \hfil Combined \hfil\cr
\noalign{\vskip 3pt}
 {\tt \smica}$\times^{\rm bc}$& $!0.31\pm0.05$& $!0.00$ & $228.7$ & $-57.9$!\cr
\noalign{\vskip 3pt\hrule\vskip 4pt}}}
\endPlancktable                    
\tablenote {{\rm a}} The bias refers to the average value of $\alpha$
determined using the corresponding FFP8.1 simulations.\par
\tablenote {{\rm b}} \nilc\ and \smica\ have smaller uncertainties compared with
\commander\ and \sevem, which follows from the naive expectation of the rms in
the polarization maps \citep[see table~1 of][]{planck2014-a11}.\par
\tablenote {{\rm c}} The $\times$ symbol denotes the cross-correlation of
half-mission 1 with half-mission 2 data.\par
\tablenote {{\rm d}} The corresponding probability to exceed is
always below $1/1000$ except for \sevem\ which turns out to be $2/1000$.\par
\endgroup
\end{table}

\begin{figure}
\centering
\includegraphics[width=9cm]{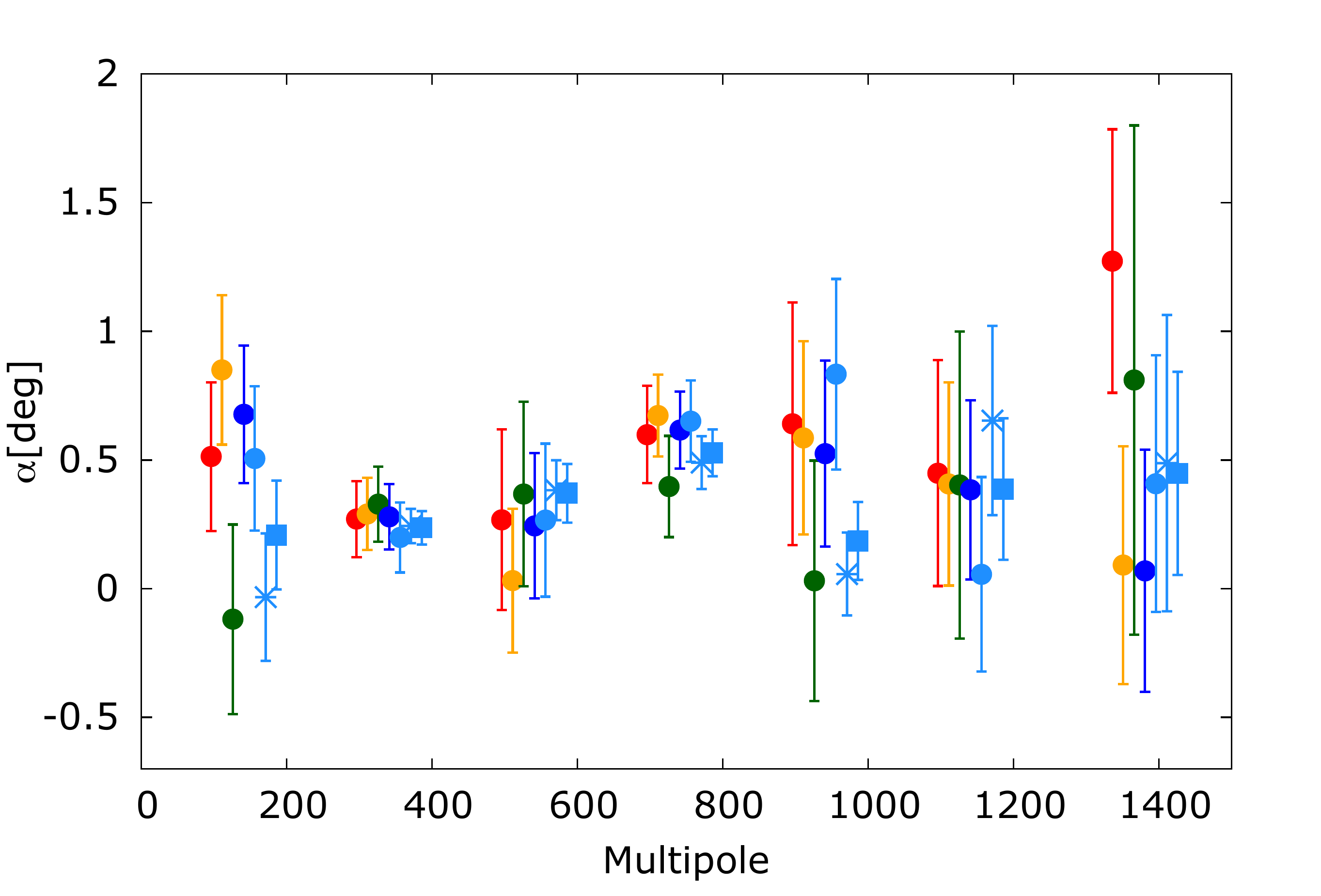}
\caption{Spectrum of $\alpha$ in degrees versus multipole. As in Fig.~\ref{one}
only ($1\,\sigma$) statistical uncertainties are shown. The colour coding is
the same as in Fig.~\ref{one}. See also footnote 9.}
\label{two}
\end{figure}

In Fig.~\ref{two} we show the dependence of $\alpha$ on angular
scale.\footnote{This should not be confused with a spectrum of the birefringence anisotropies.
As stated in Sect.~\ref{impact} in this paper we are only concerned with a
uniform rotation.}
This is
built by considering Eq.~(\ref{loglike1}) applied to $D_{\ell}^{TB}$.  Comparing
the different component-separation methods and the different ways of estimating
the spectra, we find good stability of the $\alpha$ estimates for each angular
scale.  The statistical uncertainties follow the behavior described in \citet{Gruppuso:2016nhj}.

\subsubsection{$E$--$B$}

Considering Eq.~(\ref{loglike}) for the estimator $D_{\ell}^{EB}$, defined in
Eq.~(\ref{DEB}), we have extracted the birefringence angle $\alpha$ for the
half-mission \smica\ data. The estimate obtained
for $\alpha$ is given in Table \ref{tableone} (see sixth row) and is compatible
with constraints from the other component-separation methods, as is also clear from Fig.~\ref{one}.
Fig.~\ref{two} shows the spectrum of $\alpha$ obtained in this case.  Note
that the statistical uncertainty coming from $D_{\ell}^{EB}$ is much smaller
than that obtained from $D^{TB}_{\ell}$.  This is expected, since
Eqs.~\eqref{eq:tbuncertainty} and \eqref{eq:ebuncertainty} suggests that
the $E$--$B$ correlation is able to constrain $\alpha$ better than the $T$--$B$
correlation.

\subsubsection{$T$--$B$ and $E$--$B$ combined}

The CMB power spectra from the \smica\ cross-correlations allow us to build a joint estimate
minimizing the total $\chi^2$, defined as $\chi^2(\alpha) = \chi^2_{TB}(\alpha)
+ \chi^2_{EB}(\alpha)$. We have explicitly checked with FFP8.1 simulations that there is no significant
cross-correlation between $D_{TB}$ and
$D_{EB}$\footnote{The impact of taking such a cross-correlation into account is at most
at the level of half of the statistical standard deviation.} and this in turn means that it is possible
to minimize the simple sum of $\chi^2$. Not surprisingly, it turns out that
such a combination is dominated by the $E$--$B$ correlation information.  The obtained constraint
is reported again in Table \ref{tableone} (see last row) and in Fig.~\ref{one}.  As
before, in Fig.~\ref{two} we provide the spectrum of $\alpha$ obtained from minimizing
$\chi^2$ in intervals of $\ell$.
The overall consistency of each estimate is always very good.

\section{Systematic effects}
\label{sec:systematics}

The main systematic effect that is completely degenerate with the signal from
isotropic cosmological birefringence is uncertainty in the orientation
of the PSBs used for mapmaking \citep[][]{Pagano2009}. The nature of this error is
characterized in the PCCS2 paper \citep{planck2014-a35}, as well as HFI
\citep[][]{planck2014-a08,planck2014-a09} and LFI
\citep[][]{leahy2010,planck2013-p02a,planck2013-p02d} systematics papers, and
also described in \citet{planck2014-a10}.  The present upper limit in any
global rotation of the HFI detectors is estimated to be better than $0\fdg3$;
however, the {\it relative\/} upper limit between separate PSBs is $0\fdg9$
\citep{Rosset2010}.
After converting the above numbers into standard deviations (assuming they are
approximately uniform distributions, and noting that the
relative uncertainty can be averaged over the eight PSBs used by \Planck) we
conservatively quote the total
(global and relative) 1$\,\sigma$ uncertainty as $0\fdg28$. This final error
is not exactly Gaussian, although it is close (68\,\% and 95\,\% CLs are
$0\fdg28$
and $0\fdg55$, respectively).
Given that we detect a rotation of around $0\fdg3$, we are,
therefore, unable to disentangle the signal found
in the data from the possible presence of this systematic effect.  It remains
to be seen whether or not this can be improved in a future \Planck\ release.

It might be expected that \commander\ would perform best in polarization in
terms of noise and handling of systematics \citep[based on the angular scales
probed here, see][for details]{planck2014-a11}. However, given that \commander\
uses a slightly different set of data than the other component-separation
methods and given that they all use different algorithms, we cannot make any
definitive claims as to which gives the most accurate constraint. We are also
unable to account for the apparent discrepancy at the roughly 2$\,\sigma$ level
(given the large number of comparisons performed here, this could simply be a
statistical fluke)
that the \commander\ noise estimate yields for $\alpha$ when stacking on
temperature (see Table~\ref{tab:constraints}). That being said, it is
reassuring that all component-separation methods agree at the $\simeq
1\,\sigma$ level in their constraints on $\alpha$.

In the following subsections we mention some other possible systematic effects
that might be present, but that we believe contribute negligibly to the
polarization rotation signal.

\subsection{Noise properties of polarization}
\label{sec:pnoise}

The recommendation from the \Planck\ collaboration is that any analysis performed
on polarization data should not be very sensitive to mis-characterization
of the noise. To this end cross-spectra, cross-correlation, and stacking
analyses are examples of such approaches. Our harmonic space and map space
temperature tests fulfil this criterion explicitly. It is less obvious that
stacking on $E$-mode extrema should only weakly depend on the noise properties;
however, we find this to be the case. This is because the statistics of the $E$-mode
map are encoded in the bias parameters ($\bar{b}_\nu, \bar{b}_\zeta$), which
depend on the {\it total\/} power in the map (see
Appendix~\ref{app:Emodes} for details). Therefore the bias parameters
will be accurate to the level that the statistics of the polarization data can
be determined by its two-point function. Nevertheless we will now describe to
what extent a miscalculation of the bias parameters will affect our results.

We use the {\tt MASTER} method \citep[to correct for masking,][]{Hivon2002} to
estimate the total power spectrum of the $E$-mode map in order to calculate the
bias parameters ($\bar{b}_\nu, \bar{b}_\zeta$). The noise term in
Eq.~\eqref{eq:Eqrprofile} is then given by subtracting the theoretical power
spectrum ($C^{EE}_{\ell}$) from the total power spectrum. The main effect of
noise, however, comes from the determination of the bias terms only, since most
of the discriminatory power on $\alpha$ comes from the $U_{\it r}$ stacks
(Eqs.~\ref{eq:Eurprofile} and \ref{eq:Eurmodel} do not explicitly depend on the
noise term).

The largest difference in our noise estimation when comparing between different
component-separation methods comes from \sevem\ and \smica. For these maps
$\bar{b}_\mu$ differs by approximately $20$\,\%, and $\bar{b}_\zeta$ by
$40$\,\%. Using the \sevem\ bias parameter values on the \smica\ data
(for example) leads to a roughly $1\,\sigma$ shift in the posterior of $\alpha$
(from $E$-modes). Such a discrepancy estimate is overly conservative however, because
each component-separation technique will generally produce maps with different
noise levels. If instead we scale our noise estimate by as much as 10\,\% (for
any of the individual component-separation methods) we find that $\alpha$ shifts by less
than 0.25$\,\sigma$.  We therefore conclude that for our analysis,
mis-characterization of the noise in polarization has little to no effect.

\subsection{Beam effects}
\label{sec:beams}

Because of  the differential nature of polarization measurements, any beam
mismatch or uncertainties can induce temperature-to-polarization leakage
\citep[][]{Hu2003,leahy2010}. Here we are interested in beam uncertainties that
can potentially lead to $T$--$B$ and $E$--$B$ correlations that might mimic a
non-zero $\alpha$ signal.  Due to circular symmetry, effects from differential
beam sizes or differential relative gains will not tend to produce $T$--$B$ or
$E$--$B$ correlations, whereas effects from differential pointing and
differential ellipticity will.  Differences in the noise level will also in
general cause temperature-to-polarization leakage.

We check for these effects following the approach described in \citet{planck2014-a13} and
\citet{planck2014-a15}. Note that temperature-to-polarization leakage estimates
due to bandpass mismatches between detectors have been removed from the
component-separated maps \citep[see][for details]{planck2014-a05,
planck2014-a07, planck2014-a09, planck2014-a11}; we perform a crude scan of the
parameter space in the following temperature-to-polarization leakage model
\citep[see also appendix A.6 of][]{planck2014-a10}:
\begin{align}
  C^{TE}_{\ell} &\rightarrow C^{TE}_{\ell} + \epsilon C^{TT}_{\ell}; \\
  C^{TB}_{\ell} &\rightarrow \beta C^{TT}_{\ell}; \\
  C^{EE}_{\ell} &\rightarrow C^{EE}_{\ell} + \epsilon^2 C^{TT}_{\ell} + 2
  \epsilon C^{TE}_{\ell}; \\
  C^{EB}_{\ell} &\rightarrow \epsilon \beta_{\ell} C^{TT}_{\ell} + \beta
  C^{TE}_{\ell}.
  \label{eq:ttopleakage}
\end{align}
The $\epsilon$ and $\beta$ terms are expected to be dominated by
$m=2$ and $m=4$ modes (assuming the mismatch comes from differential
ellipticity) and can be written as
\begin{align}
  \epsilon &= \epsilon_2 \ell^2 + \epsilon_4 \ell^4, \\
  \beta &= \beta_2 \ell^2 + \beta_4 \ell^4.
  \label{eq:ebterms}
\end{align}
Varying $(\epsilon_2, \beta_2)$, and $(\epsilon_4, \beta_4)$ in the range given
by $\sigma_2 = 1.25\times10^{-8}$, and $\sigma_4 = 2.7\times10^{-15}$
\citep[][]{planck2014-a13}, we find
that $\alpha$ is stable to $<0.1\,\sigma$ (this is the case for both
temperature and $E$-mode stacks).

We must stress, however, that the above temperature-to-polarization leakage model is
not completely satisfactory \citep[see section~3.4.3 and appendix C.3.5 in][for
full details]{planck2014-a13}. Nevertheless it is adequate for our purposes,
since we only wish to
demonstrate that our results remain stable to most forms of beam mismatch.

\section{Conclusions}
\label{sec:conclusions}

We have estimated the rotation, $\alpha$, of the plane of polarization of CMB photons
by using \Planck\ 2015 data.
Employing harmonic-space cross-correlations and a map-space stacking approach we
find values of
$0\fdg31$ and $0\fdg35$, respectively, for the angle $\alpha$ (using \smica\ data).
Both methods yield the same statistical uncertainty, i.e., $0\fdg05$ (68\,\% CL),
and are subject to the same systematic error of $0\fdg28$ (68\,\% CL)  due to
the uncertainty in the global and relative orientations of the PSBs.
Our results are compatible with no rotation, i.e., no parity violation, within the total error budget.
We have demonstrated that our findings are robust against two independent
analysis approaches, different component-separation methods, harmonic scales, choices in peak
thresholds, and temperature-to-polarization leakage, at better than the $1\,\sigma$ statistical
level. We have also carefully chosen our analyses to be insensitive to detailed
knowledge of the noise properties of the polarization data.
Note that the statistical and systematic error bars represent our best
knowledge of the \Planck\ data at the time of publication.\footnote{We do not recommend the use of any $TB$, and $EB$ information (either in
form of spectra, stacking or any other estimator that depends on the
cross-correlation between $T$ and $B$ modes or $E$ and $B$ modes) without
including in the analysis the uncertainty coming from the instrumental
polarization angle (and other systematic effects that might dominate the error
budget).}
Several additional effects have the potential to enlarge the error estimates.
Among the possible source of extra systematics are residuals from the
processing, which are only partially captured by the FFP8.1 simulations since
these simulations do not yet include all the details of the instruments.
In \cite{planck2014-a13} we analysed a few  end-to-end simulations from HFI,
which include more systematic effects than those contained in FFP8.1, and found
no evidence for significant influence on the results of the TT likelihood.
\cite{planck2014-a10} shows that the FFP8 simulations fail to capture most of the very low ell ($\ell<30$)  polarization systematics. Our measurement here, based on multipoles larger than $\ell=50$
should be immune to these sorts of issues, but there are not enough end-to-end
simulations available at this time to definitively prove this.
Similarly, we relied on the efficiency of component-separated maps to treat the
Galactic residuals.
This assumes that the FFP8 Galactic model correctly describes $TB$ and $EB$
induced correlation.
Comparing the estimates obtained from different component-separation methods, we expect that the latter uncertainty
is at most of the order the of $1\,\sigma$ statistical error.

\begin{figure}
\centering
\includegraphics[width=9.5cm]{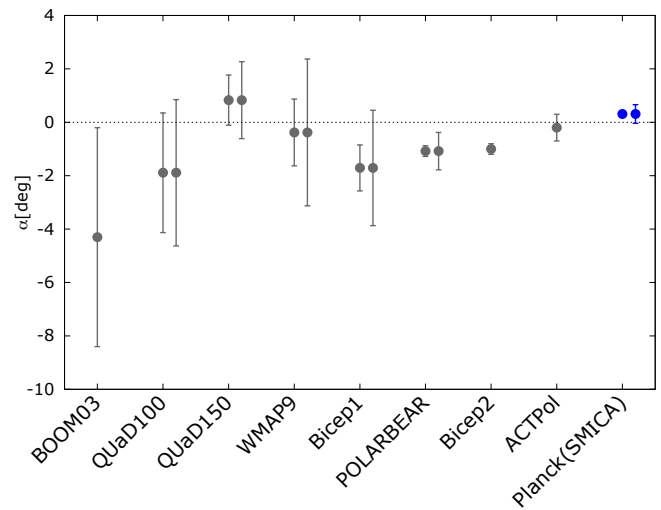}
\caption{Constraints on $\alpha$ coming from published analyses of several sets
of CMB experimental data sets (shown in grey) as reviewed in \citet{Kaufman:2014rpa}
compared with what is found in the present paper (in blue). For each experiment
the left error bars are for statistical uncertainties at 68\% CL, while
right error bars (when displayed) are obtained by summing linearly the statistical and systematic uncertainties.
The error bar of BOOM03 already contains a contribution from systematic effects.}
\label{four}
\end{figure}

In Fig.~\ref{four} we show a comparison of our estimate with the birefringence
angle estimates provided by analysis on other CMB data in which, where possible,
the total error budget is decomposed in these two parts, i.e. statistical (left point of a pair) and systematic (right).
The total error budget of our estimate is dominated by the systematic uncertainty, which is a factor of 6 larger than the statistical one.
It is clear, therefore, that future CMB polarization experiments (or a future \Planck\
release) will require a much better understating of their polarimeter orientations, since {this is appearing as the current} limiting factor of this
investigation. With a coordination of careful ground-based measurements and improved in-flight calibration on polarized sources \citep[see][for an
example of a possible effort]{Kaufman2014} we may be able to
further probe possible parity violations in the Universe.


\begin{acknowledgements}
The Planck Collaboration acknowledges the support of: ESA; CNES, and
CNRS/INSU-IN2P3-INP (France); ASI, CNR, and INAF (Italy); NASA and DoE
(USA); STFC and UKSA (UK); CSIC, MINECO, JA, and RES (Spain); Tekes, AoF,
and CSC (Finland); DLR and MPG (Germany); CSA (Canada); DTU Space
(Denmark); SER/SSO (Switzerland); RCN (Norway); SFI (Ireland);
FCT/MCTES (Portugal); ERC and PRACE (EU). A description of the Planck
Collaboration and a list of its members, indicating which technical
or scientific activities they have been involved in, can be found at
\href{http://www.cosmos.esa.int/web/planck/planck-collaboration}{\texttt{http://www.cosmos.esa.int/web/planck/planck-collaboration}}.
%
%
%
Some of the results of this paper have been derived using the {\sc HEALPix}
package \citep{Gorski2005}.
\end{acknowledgements}

\bibliographystyle{aat}
\bibliography{Planck_bib,custom,stacking_draft/birefringence}

\begin{appendix}

\section{Stacking on {$\boldsymbol E$}-mode peaks}
\label{app:Emodes}

Here we will discuss the new procedure of stacking on $E$-mode extrema. We
remind readers that the full implementation and derivation of all relevant
parameters are discussed in great detail in appendix B of \citet{Komatsu2011},
and a similar derivation is given in \citet{Contaldi2015} for stacking on $Q$ and $U$
extrema. Here we simply explain the few details required to
extend the formalism for stacking on $E$-modes.

Selecting the peaks of an underlying Gaussian field (like $T$ or $E$) leads to
a biased selection of that field. Such a bias is scale-dependent and has the
form
\begin{align}
  \delta_{\rm pk}(\vec{\hat{n}}) &= \left[b_{\nu} -
  b_{\zeta}\nabla^2\right]E(\vec{\hat{n}}).
  \label{eq:bias}
\end{align}
The scale-dependent term ($b_{\zeta}$) arises because peaks are defined by a
vanishing first derivative and the sign of the second derivative
\citep[][]{Desjacques2008}.

The bias parameters depend entirely on rms values of derivatives of the
Gaussian field, $\sigma_0,\, \sigma_1$, and $\sigma_2$ \citep[they also depend
on special functions involved in translating a 3-dimensional Gaussian random
field to the 2-dimensional case, as discussed in][]{Bond1987}. These are defined as
\begin{align}
  \sigma^2_j &\equiv \frac{1}{4\pi} \int d\vec{\hat{n}} \left(\nabla^2\right)^j
  E^2(\vec{\hat{n}}) \\
  &= \frac{1}{4\pi} \sum_{\ell} (2\ell + 1)[\ell(\ell + 1)]^j
  (C^{EE}_{\ell} + N^{EE}_{\ell}) (W^E_{\ell})^2.
  \label{eq:sigmas}
\end{align}
This is the only expression that contains the noise term $N^{EE}_{\ell}$,
which is why understanding the noise properties of the $E$-mode map is
potentially considered to be a relevant systematic effect (see
Sect.~\ref{sec:pnoise}).

When we stack $Q_{\rm r}$ or $U_{\rm r}$ on the location of $E$-mode peaks we are
explicitly computing the cross-correlation $\langle \delta_{\rm
pk}(\vec{\hat{n}}) Q_{\rm r}(\vec{\hat{n}} + \vec{\theta})\rangle$ or $\langle
\delta_{\rm pk}(\vec{\hat{n}}) U_{\rm r}(\vec{\hat{n}} + \vec{\theta})\rangle$.
Recalling that both $Q_{\rm r}$ and $U_{\rm r}$ can be written in terms of $E$ and $B$-mode
contributions \citep[][]{Zaldarriaga1997, Kamionkowski1997} and rewriting
Eq.~\eqref{eq:bias} in the flat-sky approximation we arrive at\footnote{For
brevity we henceforth drop the noise term in the expression for
$C^{EE}_{\ell}$.}
\begin{align}
  \langle \delta_{\rm pk}(\vec{\hat{n}}) Q_{\rm r}(\vec{\hat{n}} +
  \vec{\theta})\rangle = \int \frac{d^2 \vec{\ell}}{(2\pi)^2} W^E_\ell W^P_\ell
  (\bar{b}_{\nu} + \bar{b}_{\zeta} \ell^2) \notag \\
  \left\{C^{EE}_{\ell} \cos{[2(\phi - \psi)]} + C^{EB}_{\ell} \sin{[2(\phi -
  \psi)]}\right\} e^{i\vec{\ell} \cdot \vec{\theta}}, \\
  \langle \delta_{\rm pk}(\vec{\hat{n}}) U_{\rm r}(\vec{\hat{n}} +
  \vec{\theta})\rangle = \int \frac{d^2 \vec{\ell}}{(2\pi)^2} W^E_\ell W^P_\ell
  (\bar{b}_{\nu} + \bar{b}_{\zeta} \ell^2) \notag \\
  \left\{C^{EB}_{\ell} \cos{[2(\phi - \psi)]} - C^{EE}_{\ell} \sin{[2(\phi -
  \psi)]}\right\} e^{i\vec{\ell} \cdot \vec{\theta}}.
  \label{eq:stackaverage}
\end{align}
Here we have used the coordinate convention of \citet{Komatsu2011}, thus
$\vec{\ell} = (\ell \cos{\psi}, \ell \sin{\psi})$, and $\vec{\theta} = (\theta
\cos{\phi}, \theta \sin{\phi})$. We can perform the internal integration over $\psi$
using properties of Bessel functions to finally arrive at
Eqs.~\eqref{eq:Eqrprofile}--\eqref{eq:Eurprofile}, i.e.,
\begin{align}
  \langle \delta_{\rm pk}(\vec{\hat{n}}) Q_{\rm r}(\vec{\hat{n}} +
  \vec{\theta})\rangle = - \int \frac{\ell d\ell}{2\pi} & W^T_{\ell} W^P_{\ell}
  \left(\bar{b}_\nu + \bar{b}_\zeta {\ell}^2\right) \notag \\
  &C^{EE}_{\ell} J_2({\ell}\theta), \\
  \langle \delta_{\rm pk}(\vec{\hat{n}}) U_{\rm r}(\vec{\hat{n}} +
  \vec{\theta})\rangle = - \int \frac{\ell d\ell}{2\pi} & W^T_{\ell} W^P_{\ell}
  \left(\bar{b}_\nu + \bar{b}_\zeta {\ell}^2\right) \notag \\
  &C^{EB}_{\ell} J_2({\ell}\theta).
  \label{eq:appprofiles}
\end{align}

These angular profiles could have been derived in a more heuristic way by
realizing that an $E$-mode map has the same statistical properties as a
temperature map, differing only in its power spectrum. Thus we could have gone
from Eqs.~\eqref{eq:qrprofile}--\eqref{eq:urprofile} to
Eqs.~\eqref{eq:Eqrprofile}--\eqref{eq:Eurprofile} by simply making the
replacement $T \rightarrow E$.

\end{appendix}

\raggedright

\end{document}